\documentclass[journal]{IEEEtran}

\ifCLASSINFOpdf
\else
\fi
\usepackage{graphics,graphicx,epsf,psfrag, subfigure,epsfig, times,amsmath,amssymb,color,kbordermatrix}
\usepackage{url}
\usepackage{verbatim} 
\usepackage{multicol}
\usepackage{array}
\usepackage{margins}
\usepackage{psfrag}

\usepackage{amsthm}
\usepackage{amsmath, amssymb}
\usepackage{amsfonts}

\newcommand{\rvx}{\mathsf{x}}

\newcommand{\rvs}{\mathsf{s}}

\newcommand{\cC}{{\mathcal C}}
\newcommand{\cT}{{\mathcal T}}
\newcommand{\g}{\gamma}
\newcommand{\al}{\alpha}
\newcommand{\cp}{\check{p}}

\newcommand{\bp}{\mathbf{p}}
\newcommand{\bs}{\mathbf{s}}
\newcommand{\bq}{\mathbf{q}}
\newcommand{\bd}{\mathbf{b}}

\newcommand{\CapTau}{\mathcal{T}}
\newcommand{\cB}{{\mathcal B}}
\newcommand{\cA}{{\mathcal A}}
\newcommand{\obvec}{\bar{\bvec}}
\newcommand{\odvec}{\bar{\dvec}}
\newcommand{\op}{\bar{p}}
\newcommand{\osvec}{\bar{\svec}}
\newcommand{\mat}[1]{\begin{bmatrix} #1 \end{bmatrix}}
\newcommand{\Gmat}{\mathbf{G}}
\newcommand{\Imat}{\mathbf{I}}
\newcommand{\Hmat}{\mathbf{H}}
\newcommand{\Zeromat}{\mathbf{0}}
\newcommand{\uvec}{\mathbf{u}}
\newcommand{\nvec}{\mathbf{n}}
\newcommand{\bvec}{\mathbf{b}}
\newcommand{\cvec}{\mathbf{c}}
\newcommand{\pvec}{\mathbf{p}}
\newcommand{\svec}{\mathbf{s}}
\newcommand{\xvec}{\mathbf{x}}

\newcommand{\dvec}{\mathbf{d}}
\newcommand{\rvec}{\mathbf{r}}
\newcommand{\Fq}{\mathbb{F}_Q}

\newtheorem{thm}{Theorem}
\newtheorem{lem}{Lemma}
\newtheorem{proposition}{Proposition}

\newtheorem{definition}{Definition}

\begin{document}

\title{Diversity Embedded Streaming Erasure Codes (DE-SCo): Constructions and Optimality}

\author{Ahmed Badr, Ashish Khisti,~\IEEEmembership{Member IEEE}, and Emin Martinian\thanks{Manuscript received July, 2003; revised December, 2010. Ahmed Badr and Ashish Khisti are with the University of Toronto, Toronto, ON, Canada email: \{akhisti, abadr\}@comm.utoronto.ca. Emin~Martinian is with the Massachusetts Institute of Technology (MIT), Cambridge, MA, 02139, USA. Email: emin@alum.mit.edu}}


\maketitle

\begin{abstract}
Streaming erasure codes encode a source stream to guarantee that each source symbol is recovered within a fixed delay at the receiver over a burst-erasure channel. This paper introduces  \emph{diversity embedded streaming erasure codes} (DE-SCo), that provide a flexible tradeoff between the channel quality and receiver delay. When the channel conditions are good, the source stream is recovered with a  low delay, whereas when the channel conditions are poor the source stream is still recovered, albeit with a larger delay.  Information theoretic analysis of the underlying burst-erasure broadcast channel reveals that  DE-SCo  achieve the minimum possible delay for the weaker user, without sacrificing the performance of the stronger user.   Our constructions are explicit, incur polynomial time encoding and decoding complexity and outperform random linear codes  over bursty erasure channels. 
\end{abstract}

\IEEEpeerreviewmaketitle

\begin{IEEEkeywords}
Low Delay, Streaming Erasure Correction Codes,  Burst Erasure Channel, Broadcast Channel, Network Information Theory, Delay Constrained Coding, Application Layer Error Correction  
\end{IEEEkeywords}

\vspace{-1em}
\section{Introduction}
\IEEEPARstart{F}{orward} error correction codes  designed for streaming sources require that (a) the channel input stream be produced sequentially from the source stream (b) the decoder sequentially reconstruct the source stream as it observes the channel output. In contrast, traditional error correction codes such as maximum distance separable (MDS) codes map blocks of data to a codeword and the decoder waits until the entire codeword is received before the source data can be reproduced. Rateless codes such as the digital fountain codes are not ideally suited for streaming sources. First they require that the entire source data be available before the output stream is reproduced. Secondly they provide no guarantees on the sequential reconstruction of the source stream.  Nevertheless there has been a significant interest in adapting such constructions for streaming applications see e.g.,~\cite{ftn1, ftn2, ftn3, ftn4, ftn5, ftn6}.  

In~\cite[Chapter 8]{Martinian_Thesis} a class of systematic time-invariant convolutional codes~ \emph{streaming erasure codes} (SCo) are proposed for the burst erasure channel.  The encoder observes a semi-infinite source stream and maps it to a coded output stream of rate $R$. The channel considered is a burst-erasure channel --- starting at an arbitrary time, it introduces an erasure-burst of maximum length $B$. The decoder is required to reconstruct each source symbol with a maximum delay $T$. A fundamental relationship between $R$, $B$ and $T$ is established and  SCo codes are constructed that achieve this tradeoff. We emphasize that the parity check symbols in these  constructions involve a careful combination of source symbols. In particular, random linear combinations, popularly used in e.g., network coding, do not attain the optimal performance.

The SCo framework however requires that the value of $B$ and $T$ be known apriori. In practice this forces a conservative design i.e., we design the code for the worst case $B$ thereby incurring a higher overhead (or a larger delay) even when the channel is relatively good. Moreover there is often a flexibility in the  allowable delay. Techniques such as adaptive media playback~\cite{Girod} have been designed to tune the play-out rate as a function of the received buffer size to deal with a temporary increase in delay. Hence it is  not desirable to have to fix $T$ during the design stage either.

We introduce a class of streaming codes that do not commit apriori to a specific delay. Instead they realize a delay that depends on  the channel conditions. At an information theoretic level, our setup extends the point-to-point link in~\cite{Martinian_Thesis} to a multicast model --- there is one source stream and two receivers. The channel for each receiver introduces an erasure-burst of length $B_i$ and each receiver can tolerate a delay of $T_i$ for $i=1,2$. We investigate diversity embedded streaming erasure codes (DE-SCo). These codes modify a single user SCo such that the resulting code can support a second user, whose channel introduces a larger erasure-burst, without sacrificing the performance of the first user.
Our construction  embeds new parity checks in an SCo code in  a manner such that (a) no interference is caused to the stronger (and low delay) user and (b) the weaker user can use some of the parity checks of the stronger user as side information to recover part of the source symbols. DE-SCo constructions  outperform  baseline schemes that  simply concatenate the single user SCo for the two users. An information theoretic converse establishes that DE-SCo achieves the minimum possible delay for the weaker receiver without sacrificing the performance of the stronger user. Finally all our code constructions can be encoded and decoded with a polynomial time complexity in $T$ and $B$.

In recent works,~\cite{Sahai,JayKumar,Yao} study the low-delay codes with feedback, the compression of streaming sources is studied in~\cite{Draper} while a comparison of block and streaming codes for low delay systems is provided in~\cite{Zhi}.

\section{System Model}
\label{Background}
\begin{figure}
\centering
\resizebox{\columnwidth}{!}{\includegraphics[scale=1, trim = 15mm 250mm 10mm 10mm, clip]{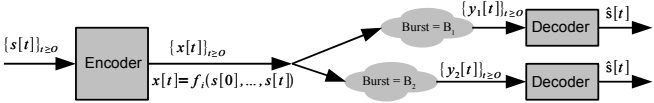}}
\caption{The source stream $\{\rvs[i]\}$ is causally mapped into an output stream $\{\rvx[i]\}$. Both the receivers observe these symbols via their channels. The channel introduces an erasure-burst of length $B_i$, and each receiver tolerates a delay of $T_i$, for $i=1,2$.\label{ProblemFormulationFigure}}
\vspace{-1 em}
\end{figure}
The transmitter encodes a stream of source symbols $\{ s[t] \}_{t \geq 0}$ intended to be received at two receivers as shown in Fig.~\ref{ProblemFormulationFigure}. The channel symbols $\{ x[t] \}_{t \geq 0}$ are produced causally from the source stream,

\vspace{-1em}

\begin{equation}
\label{Code_Function}
x[t] = f_t (s[0],\dots,s[t]).
\end{equation}
The channel of receiver $i$ introduces an erasure-burst of length $B_i$ i.e., the channel output at receiver $i$ at time $t$ is given by

\vspace{-1em}

\begin{equation}
\label{Channel_Function_Multi}
y_i[t] = 
\left\{
\begin{array}{ll}
\star & t \in [j_i,j_i + B_i - 1] \\
x[t] & \text{otherwise}
\end{array}
\right.
\end{equation}
for $i=1,2$ and for some $j_i \ge 0$. Furthermore, user $i$ tolerates a delay of $T_i$, i.e., there exists a sequence of decoding functions $\gamma_{1t}(.)$ and $\gamma_{2t}(.)$ such that
\begin{equation}
\label{Decoders}
\hat{s}_i[t] = \gamma_{it} (y_i[0],y_i[1],\dots,y_i[t+T_i]), \qquad i=1,2,
\end{equation}
and $\text{Pr} (s_i[t] \neq \hat{s}_i[t]) = 0, \; \; \; \;	\forall t \geq 0 \; \;$.

The source stream is an i.i.d.\ sequence and  we assume that each symbol  is sampled from a distribution $p_\rvs(\cdot)$ over the finite field $\Fq^T$. The rate of the multicast code is defined as ratio of the entropy of the source symbol to the (marginal) entropy of each channel symbol i.e., $R = H(\rvs)/H(\rvx)$. An \emph{optimal multicast streaming erasure code (MU-SCo)} achieves the maximum rate for a given choice of $(B_i, T_i)$. Of particular interest is the following subclass.
\begin{definition}[Diversity Embedded Streaming Erasure Codes (DE-SCo)] 
\label{defn:DE-SCo}
Consider the multicast model in Fig.~\ref{ProblemFormulationFigure} where the channels of the two receivers introduce an erasure burst of lengths $B_1$ and $B_2$ respectively with $B_1 < B_2$.  A DE-SCo is a rate $R =\frac{T_1}{T_1+B_1}$ MU-SCo construction that achieves a delay $T_1$ at receiver $1$ and supports  receiver $2$ with delay $T_2$. An optimal DE-SCo minimizes the delay $T_2$ at receiver $2$ for given values of $B_1$, $T_1$ and $B_2$. \end{definition}


Note that our model only considers a single erasure burst on each channel. As is the case with (single user) SCo, our constructions correct multiple erasure-bursts  separated sufficiently apart. Also we only consider the erasure channel model. It naturally arises when these codes are implemented in application layer multimedia encoding. More general channel models can be transformed into an erasure model by applying an appropriate inner code~\cite[Chapter 7]{Martinian_Thesis}.

\section{Background: Streaming Codes (SCo) }
\label{Background_Real}
Streaming burst-erasure codes developed in \cite{Martinian_Thesis} and \cite{Martinian_Trott} are single user codes for the model in the previous section. They correct an erasure burst of length $B$ with a delay of $T$ symbols and achieve the largest possible rate
\begin{equation}
\label{SingleUser_Capacity}
C = 
\left\{
\begin{array}{ll}
\frac{T}{T+B} & T \geq B \\
0 & \text{otherwise.}
\end{array}
\right.
\end{equation}

\subsection{Construction}
\label{subsec:SCo_Encoding}
The construction in~\cite{Martinian_Thesis} is described in three steps. 
\begin{enumerate}
\item Create $(T,T-B)$ Burst Erasure Block Code (BEBC) \\
The construction begins with a systematic generator matrix $\Gmat$ for a $(T,T-B)$ Burst Erasure Block Code (BEBC) over a finite field $\Fq$, without regard to decoding delay. The code must also correct ``end-around" bursts. Recall that any $(n',k')$ cyclic code corrects burst erasures of length $n' - k'$. Since the matrix $\Gmat$ is systematic we can express it in the form\vspace{-1em}

\begin{equation}
\vspace{-0.5em}\Gmat = \kbordermatrix{  & T- B & B \\ (T- B) & \Imat & \Hmat}\label{eq:Gmat}\end{equation}
where $\Imat$ denotes the identity matrix and $\Hmat$ is a $(T-B)\times B$ matrix.

\item Create $(B+T,T)$ Low-Delay Burst Erasure Block Code (LD-BEBC) \\
The LD-BEBC code maps a vector of $T$ information symbols $\bvec \in \Fq^T$ to a systematic codeword $\cvec \in \Fq^{T+B}$ as follows. We first split $\bvec$ into two sub-vectors of lengths $B$ and $T-B$

\vspace{-1em}

\begin{equation}
\bvec = \kbordermatrix{& B & T-B \\ & \uvec & \nvec},\label{eq:Urgent_NonUrgent_Split}\end{equation}
and the resulting codeword is\footnote{All addition in this paper is defined over $\Fq$ or its extension field.}\begin{align}
\cvec &= \mat{\uvec & \nvec} \cdot \mat{
\Imat_{B \times B} & \Zeromat_{B \times (T-B)} & \Imat_{B \times B} \\
\Zeromat_{(T-B) \times B} & \Imat_{(T-B) \times (T-B)} & \Hmat} \label{eq:LD_BEBC1}\\
&=\mat{\uvec & \nvec & \uvec + \nvec\cdot \Hmat}
=\mat{\bvec & \rvec} \label{eq:LD_BEBC2}
 \end{align}
 where we have used~\eqref{eq:Urgent_NonUrgent_Split} and introduced $\rvec = \uvec + \nvec \cdot \Hmat$ to denote the parity check symbols in $\cvec$ in the last step.

The codeword $\cvec$ has the property that it is able to correct any erasure burst of length $B$ with a delay of at-most $T$ symbols. If we express $\bvec = (b_0,\ldots, b_{T-1})$ then, for any erasure-burst of length $B$, $b_0$ is recovered at time $T$, $b_1$ at time $(T+1)$ and $b_{B-1}$ at time $(T+B-1)$. The remaining symbols $b_B \ldots, b_{T-1}$ are all recovered at the end of the block. 

The information symbols in vector $\uvec = (b_0, \ldots, b_{B-1})$ are referred to as \emph{urgent} symbols whereas the symbols in vector $\nvec = (b_B,\ldots, b_{T-1})$ are referred to as non-urgent symbols.

\item Diagonal Interleaving \\
The final step is to construct a streaming code (SCo) from the  LD-BEBC code in step 2. Recall that the SCo specified a mapping between the symbols $s[t]$ of the incoming source stream to the symbols $x[t]$ of the channel input stream. This mapping is of the form
\begin{equation}
\label{eq:ScoMap}
s[t]=\mat{s_0[t] \\\vdots \\s_{T-1}[t]} \in \Fq^T, \qquad x[t] = \mat{s_0[t] \\ \vdots \\ s_{T-1}[t] \\ p_0[t] \\\vdots \\ p_{B-1}[t]}\in \Fq^{T+B}
\end{equation}
i.e., we split each source symbol $s[t] \in \Fq^T$ into $T$ equal sized sub-symbols over $\Fq$ and then append $B$ parity check sub-symbols over $\Fq$. Thus we have that $x[t] \in \Fq^{T+B}$. The parity check sub-symbols $p_0[t],\ldots, p_{B-1}[t]$ are constructed through a diagonal interleaving technique described below.
 
An information vector $\bvec_t$ in~\eqref{eq:Urgent_NonUrgent_Split} is constructed by collecting sub-symbols along the diagonal of the sub-streams i.e.,\begin{equation}
\bvec_t = (s_0[t],s_1[t+1],\!\ldots,\!s_{T-1}[t+T-1]).\label{eq:diagonalMap_bvec}
\end{equation} The corresponding codeword $\cvec_t = (\bvec_t,\rvec_t)$ is then constructed according to~\eqref{eq:LD_BEBC1}. The resulting parity check sub-symbols in $\rvec[t]$ are then appended diagonally to the source stream to produce the channel input stream i.e., \begin{equation}
(p_0[t+T], \ldots, p_{B-1}[t+T+B-1])=(r_0[t],\ldots, r_{B-1}[t]) \label{eq:diagonalMap_rvec}
\end{equation} 

Notice that the operations in~\eqref{eq:ScoMap},~\eqref{eq:diagonalMap_bvec} and~\eqref{eq:diagonalMap_rvec} construct a codeword diagonally across the incoming source sub-streams as illustrated in Table.~\ref{Code23}. A \emph{diagonal codeword} is of the form \begin{align}
\label{eq:diagonalMap_dvec2}
\dvec_t \!&=& \!(\!s_0[t],\ldots, s_{T-1}[t+T-1], p_0[t+T],\\ \nonumber & &\!\ldots,\!p_{B-1}[t+T+B-1]).
\end{align}

The SCo code is  a  time-invariant convolutional code~\cite{Forney}. The inputs to the convolutional code are source symbols $\svec \in \Fq^T$, while the outputs are channel symbols $\xvec \in \Fq^{T+B}$. We emphasize that the actual transmitted symbol is given in~\eqref{eq:ScoMap}. The diagonal codeword~\eqref{eq:diagonalMap_dvec2} above  simply maps the LD-BEBC to a SCo.


\end{enumerate}

\begin{table*}[t]
\begin{tabular}{|c|c|c|c|c|c|}
\hline
\fbox{$s_0[i-1]$} & $s_0[i]$ & $s_0[i+1]$ & $s_0[i+2]$ & $s_0[i+3]$ & $s_0[i+4]$ \\
$s_1[i-1]$ & \fbox{$s_1[i]$} & $s_1[i+1]$ & $s_1[i+2]$ & $s_1[i+3]$ & $s_1[i+4]$ \\
$s_2[i-1]$ & $s_2[i]$ & \fbox{$s_2[i+1]$} & $s_2[i+2]$ & $s_2[i+3]$ & $s_2[i+4]$ \\\hline
$s_0[i-4] + s_2[i-2]$ & $s_0[i-3] + s_2[i-1]$ & $s_0[i-2] + s_2[i]$ & \fbox{$s_0[i-1] + s_2[i+1]$} & $s_0[i] + s_2[i+2]$ & $s_0[i+1] + s_2[i+3]$ \\
$s_1[i-4] + s_2[i-3]$ & $s_1[i-3] + s_2[i-2]$ & $s_1[i-2] + s_2[i-1]$ & $s_1[i-1] + s_2[i]$ & \fbox{$s_1[i] + s_2[i+1]$} & $s_1[i+1] + s_2[i+2]$\\\hline
\end{tabular}
\caption{A (2,3) Single User SCo Code Construction is given where each source symbol $\svec[.]$ is divided into three sub-symbols $s_0[.]$, $s_1[.]$ and $s_2[.]$ and a $(5,3)$ LD-BEBC code is then applied across the diagonal to generate two parity check sub-symbols generating a rate $3/5$ code. Each column corresponds to one channel symbol.}
\label{Code23}
\end{table*}

\subsection{Decoding of SCo Codes}
\label{subsec:SCo_Decoding}
The structure of the diagonal codeword~\eqref{eq:diagonalMap_dvec2} is also important in decoding. Suppose that symbols $x[t],\ldots, x[t+B-1]$ are erased.  It can be readily verified that there are no more than $B$ erasures in each diagonal codeword $\{ \dvec_t \}$ (c.f. \eqref{eq:diagonalMap_dvec2}).  Since each codeword is a $(T+B, T)$ LD-BEBC, it recovers each erased symbol with a delay of no more than $T$ symbols. This in turn implies that all erased symbols are recovered.

\subsection{Example: (2,3) SCo Code} 
Suppose we wish to construct a code capable of correcting any symbol burst erasure of length $B=2$ with delay $T=3$. A  LD-BEBC \eqref{eq:LD_BEBC1} for these parameters is 
\begin{equation}
\label{eq:LDBEBC31}
\cvec = (b_0,b_1,b_2,b_0 + b_2, b_1 + b_2).
\end{equation}To construct the SCo code, we divide the source symbols  into $T=3$ sub-symbols. The diagonal codeword~\eqref{eq:diagonalMap_dvec2} is of the form\begin{equation}
\label{eq:dvec_23}
\!\dvec_t\!=\!(\!s_0[t],s_1[t+1], s_2[t+2], s_0[t]+s_2[t+2],s_1[t+1]+s_2[t+2]\!)\!\end{equation}
and the channel input $x(t)$ is given by
\small
\begin{equation}
x[t] = \mat{s_0[t] , s_1[t] , s_2[t] , s_0[t-3] + s_2[t-1] , s_1[t-3] + s_2[t-2]}^\dagger.
\end{equation}
\normalsize
The resulting channel input stream is illustrated in Table.~\ref{Code23}. Note that the rate of this code is $3/5$ as it introduces two parity check sub-symbols for each three source sub-symbols. It can be easily verified that this code corrects a burst erasure of length $2$ with a worst-case time delay $3$.

\section{SCo Properties}
\begin{figure}
\centering
\resizebox{\columnwidth}{!}{\includegraphics[scale=0.65]{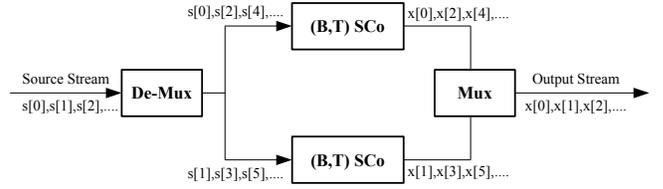}}
		\caption{A vertical interleaving approach to construct a $(2B,2T)$ SCo code from a $(B,T)$ SCo code.
		\label{Vertical_Interleaving}}
\end{figure}
\label{subsec:SCoProperties}
In this section we describe some additional properties of SCo codes that will be useful in the DE-SCo construction.

\subsection{Vertical Interleaving for $(\al B, \al T)$ SCo}
\label{subsec:Vertical_Interleaving} Suppose $\al \ge 2$ is an integer and we need to construct a SCo code with parameters $(\al B, \al T)$. The scheme described in section~\ref{subsec:SCo_Encoding} requires us to split each source symbol into $\al T$ sub-symbols. However we can take advantage of the multiplicity factor $\al$ and simply construct the $(\al B, \al T)$ SCo code from the $(B,T)$ SCo code via vertical interleaving of step $\al$.

Fig.~\ref{Vertical_Interleaving} illustrates this approach for constructing a $(2B,2T)$ SCo from a $(B,T)$ SCo. We split the incoming source stream into two disjoint sub-streams; one consisting of source symbols at even time slots and the other consisting of symbols at odd time slots. We apply a $(B,T)$ SCo on the first sub stream to produce channel symbols at even time slots. Likewise we apply a $(B,T)$ SCo on the second sub stream to produce channel symbols at odd time slots. Since a burst of length $2B$ introduces $B$ erasures on either sub-streams, each of the $(B,T)$ code suffices to recover from these erasures. Further each erased symbol is recovered with a delay of $T$ symbols on its individual sub stream, which corresponds to an overall delay of $2T$ symbols.

More generally we split each source symbol into $T$ sub-symbols. The information vector~$\bvec_t$ is modified 
from~\eqref{eq:diagonalMap_bvec} as\begin{equation}
\bvec_t = (s_0[t],s_1[t+\al],\!\ldots,\!s_{T-1}[t+(T-1)\al]).\label{eq:diagonalMap_bvec_al}
\end{equation}
The resulting codeword $\cvec[t]$ of the LD-BEBC is then mapped to a \emph{diagonal codeword} by introducing a step-size of $\al$ in~\eqref{eq:diagonalMap_dvec2} i.e.,
\begin{multline}
\label{eq:diagonalMap_dvec_al}
\dvec_t = (s_0[t],s_1[t+\al],\ldots, s_{T-1}[t+(T-1)\al], \\
p_0[t+T\al],\ldots, p_{B-1}[t+(T+B-1)\al]).
\end{multline}
As in the case of $\al=2$, the decoding proceeds by splitting the source stream into $\al$ sub-streams and applying the decoder for $(B,T)$ SCo on each of the sub-streams. This guarantees that each symbol is recovered with a delay of $\al T$ on the original stream.

\subsection{Memory in Channel Input Stream $\{x[t]\}$}
\label{subsec:Memory}
While the definition of SCo allows the channel input symbol $x[t]$ to depend on an arbitrary number of source symbols, the construction limits the memory of symbol $x[t]$ to previous $T$ symbols i.e.,
\begin{equation}
\label{eq:ScoMemory}
x[t] = f(s[t],s[t-1],\ldots, s[t-T]).
\end{equation}Furthermore a closer look at the parity check sub-symbols~\eqref{eq:ScoMap} of $x[t]$ reveals that the parity checks $p_0[t],\ldots, p_{B-1}[t]$ constructed from the LD-BEBC in~\eqref{eq:LD_BEBC2} have the form
\begin{multline}
p_j[t] = s_j[t-T] + h_j(s_B[t-j-T+B],\ldots, s_{T-1}[t-j-1]), \\j=0,\ldots, B-1,
\label{eq:ScoParity}
\end{multline}
where $h_j(\cdot)$ denotes a linear combination arising from the LD-BEBC code~\eqref{eq:LD_BEBC2} when applied along the main diagonal.

\subsection{Urgent and Non-Urgent Sub-Symbols}
\label{subsec:UrgentNonUrgent}
In the construction of LD-BEBC codes we split the information vector $\bvec$ into urgent and non-urgent sub-symbols~\eqref{eq:Urgent_NonUrgent_Split}.  The mapping of source sub-symbols to information vector~\eqref{eq:diagonalMap_bvec} then implies that the sub-symbols $s_0,\ldots, s_{B-1}$ are the urgent sub-symbols in the source stream whereas the sub-symbols $s_B,\ldots, s_{T-1}$ are the non-urgent sub-symbols. We will denote these by
\begin{align}
\label{eq:UrgentNonUrgent_Defn}
\svec^U[t] = (s_0[t],\ldots, s_{B-1}[t]), \nonumber \\ \svec^N[t] = (s_B[t],\ldots, s_{T-1}[t]).
\end{align}
The urgent and non-urgent sub-symbols are combined into a parity check sub-symbol as illustrated in~\eqref{eq:ScoParity}.
The following observation is useful in the construction of DE-SCo.
\begin{proposition}
 \label{prop:UrgentNonUrgent_Decoding}
 Suppose that the sequence of channel symbols $x[i-B],\ldots, x[i-1]$ are erased by the burst-erasure channel. Then \begin{enumerate}
\item All sub-symbols in $\svec^N[i-B],\ldots, \svec^N[i-1]$ are obtained from the parity checks $\pvec[i],\ldots,\pvec[i+T-B-1]$.
\item The sub-symbols in $\svec^U[j]$ for $i-B\le j < i$ are recovered at time $j+T$ from parity check $\pvec[j+T]$ and the previously recovered non-urgent sub-symbols.
 \end{enumerate}
\end{proposition}
The proof follows via~\eqref{eq:ScoMemory},~\eqref{eq:ScoParity} and will be omitted due to space constraints.


\subsection{Off-Diagonal Interleaving}
The constructions in section~\ref{subsec:SCo_Encoding} involve interleaving along the main diagonal of the source stream (c.f.~\eqref{eq:diagonalMap_dvec2},\eqref{eq:diagonalMap_bvec}). An analogous construction of the $(B,T)$ code along the off diagonal results in
\begin{align}
\obvec_t &=  \left(s_0[t], s_1[t-1],\ldots, s_{T-1}[t- (T-1)]\right)\\
\odvec_t &=  (s_{T-1}[t- (T-1),\ldots, s_1[t-1],s_0[t], \nonumber\\ &\quad\quad \op_0[t+1],\ldots, \op_{B-1}[t+B]])
\end{align}
and the parity checks $\op_j$ are given by
\begin{align}
\label{eq:ScoParity_Oppossite}
\op_j[t] = s_{T-j-1}[t-T] + h_j(s_{T-B-1}[t-j-T+B], \\ \nonumber \ldots, s_{0}[t-j-1]), \quad \quad \quad j=0,\ldots, B-1,
\end{align}
when applied along the opposite diagonal. Finally off-diagonal interleaving also satisfies Prop.~\ref{prop:UrgentNonUrgent_Decoding} provided with appropriate modifications in the definitions of urgent and non-urgent sub-symbols
\begin{align}
\label{eq:UrgentNonUrgent_Defn_off}
\osvec^U[t] = (s_{T-1}[t],\ldots, s_{T-B}[t]), \nonumber \\ \osvec^N[t] = (s_{T-B-1}[t],\ldots, s_{0}[t]).
\end{align}

\section{Example}
\label{sec:Example}
\begin{table*}[t]
\centering
\subfigure[SCo Construction for $(B,T) = (1,2)$]{
\label{Code1224_a}
\begin{tabular}{|c|c|c|c|c|c|}
\hline
\fbox{$s_0[i-1]$} & $s_0[i]$ & $s_0[i+1]$ & $s_0[i+2]$ & $s_0[i+3]$ & $s_0[i+4]$ \\
$s_1[i-1]$ & \fbox{$s_1[i]$} & $s_1[i+1]$ & $s_1[i+2]$ & $s_1[i+3]$ & $s_1[i+4]$ \\
\hline
$s_0[i-3] + s_1[i-2]$ & $s_0[i-2] + s_1[i-1]$ & \fbox{$s_0[i-1] + s_1[i]$} & $s_0[i] + s_1[i+1]$ & $s_0[i+1] + s_1[i+2]$ & $s_0[i+2] + s_1[i+3]$\\
\hline
\end{tabular}
}
\subfigure[SCo Construction for $(B,T) = (2,4)$]{
\begin{tabular}{|c|c|c|c|c|c|}
\hline
\fboxrule=1pt
\fbox{$s_0[i-1]$} & $s_0[i]$ & $s_0[i+1]$ & $s_0[i+2]$ & $s_0[i+3]$ & $s_0[i+4]$ \\
$s_1[i-1]$ & $s_1[i]$ & \fboxrule=1pt \fbox{$s_1[i+1]$} & $s_1[i+2]$ & $s_1[i+3]$ & $s_1[i+4]$ \\
\hline
$s_0[i-5] + s_1[i-3]$ & $s_0[i-4] + s_1[i-2]$ & $s_0[i-3] + s_1[i-1]$ & $s_0[i-2] + s_1[i]$ & \fboxrule=1pt \fbox{$s_0[i-1] + s_1[i+1]$} & $s_0[i] + s_1[i+2]$\\
\hline
\end{tabular}
\label{Code1224_b}
}
\caption{Single user SCo constructions are shown in the upper two figures. Note that the $(1,2)$ SCo code recovers a single erasure with a delay $T=2$ but cannot recover from $B=2$. The $(2,4)$ SCo code recovers from $B=2$ with a delay of $T=4$ but does not incur a smaller delay when $B=1$.}
\label{Code1224}
\end{table*}

\normalsize

We first highlight our results via a numerical example: $(B_1,T_1)= (1,2)$ and $(B_2,T_2) = (2,4)$. Single user SCo constructions from~\cite{Martinian_Thesis,Martinian_Trott} for both users are illustrated in  Table~\ref{Code1224}(a) and~\ref{Code1224}(b) respectively. In each case, the source symbol $\rvs[i]$ is split into two  sub-symbols $(\rvs_0[i],\rvs_1[i])$ and the channel symbol $\rvx[i]$ is obtained by concatenating the source symbol $\rvs[i]$ with a parity check symbol $p[i]$. In the $(1,2)$ SCo construction, parity check symbol $p^{\rm{I}}[i] = s_1[i-1]+ s_0[i-2]$ is generated by combining the source sub-symbols diagonally across the source stream as illustrated with the rectangular boxes. For the $(B,T) = (2,4)$, the choice $p^{\rm{II}}[i] = s_1[i-2]+ s_0[i-4]$ is similar to the $(1,2)$ SCo, except that an interleaving of step of size $2$ is applied before the parity checks are produced.  Note that both these codes are single user codes and do not adapt to channel conditions.

\begin{table*}[t]
\centering
\subfigure[IA-SCo Code Construction for $(B_1,T_1) = (1,2)$ and $(B_2,T_2) = (2,6)$]{
\begin{tabular}{|c|c|c|c|c|c|}
\hline
$s_0[i-1]$ & \fbox{$s_0[i]$} & $s_0[i+1]$ & $s_0[i+2]$ & $s_0[i+3]$ & $s_0[i+4]$ \\
$s_1[i-1]$ & \fboxrule=1pt \fbox{$s_1[i]$} & \fbox{$s_1[i+1]$} & $s_1[i+2]$ & $s_1[i+3]$ & $s_1[i+4]$ \\
\hline
$s_0[i-3] + s_1[i-2]$ & $s_0[i-2] + s_1[i-1]$ & $s_0[i-1] + s_1[i]$ & \fbox{$s_0[i] + s_1[i+1]$} & $s_0[i+1] + s_1[i+2]$ & $s_0[i+2] + s_1[i+3]$ \\
$+$ & $+$ & $+$ & $+$ & $+$ & $+$ \\
$s_0[i-7] + s_1[i-5]$ & $s_0[i-6] + s_1[i-4]$  & $s_0[i-5] + s_1[i-3]$ & $s_0[i-4] + s_1[i-2]$ & $s_0[i-3] + s_1[i-1]$ & \fboxrule=1pt \fbox{$s_0[i-2] + s_1[i]$} \\
\hline
\end{tabular}}
\subfigure[DE-SCo Code Construction for $(B_1,T_1) = (1,2)$ and $(B_2,T_2) = (2,5)$]{\begin{tabular}{|c|c|c|c|c|c|}
\hline
\fbox{$s_0[i-1]$} & \fboxrule=1pt \fbox{$s_0[i]$} & $s_0[i+1]$ & $s_0[i+2]$ & $s_0[i+3]$ & $s_0[i+4]$ \\
\fboxrule=1pt \fbox{$s_1[i-1]$} & \fbox{$s_1[i]$} & $s_1[i+1]$ & $s_1[i+2]$ & $s_1[i+3]$ & $s_1[i+4]$ \\
\hline
$s_0[i-3] + s_1[i-2]$ & $s_0[i-2] + s_1[i-1]$ & \fbox{$s_0[i-1] + s_1[i]$} & $s_0[i] + s_1[i+1]$ & $s_0[i+1] + s_1[i+2]$ & $s_0[i+2] + s_1[i+3]$ \\
$+$ & $+$ & $+$ & $+$ & $+$ & $+$ \\
$s_1[i-6] + s_0[i-5]$ & $s_1[i-5] + s_0[i-4]$  & $s_1[i-4] + s_0[i-3]$ & $s_1[i-3] + s_0[i-2]$ & $s_1[i-2] + s_0[i-1]$ & \fboxrule=1pt \fbox{$s_1[i-1] + s_0[i]$}\\
\hline
\end{tabular}}
\caption{Rate $2/3$ code constructions that satisfy user 1 with $(B_1,T_1) = (1,2)$ and user $2$ with $B_2 = 2$.}
\label{Code1225}
\end{table*}

In Table~\ref{Code1225}(a) we illustrate a construction that achieves a rate $2/3$ and $(B_1,T_1)=(1,2)$ and still enables user 2 to recover the entire stream with a delay of $T_2=6$. It is obtained by shifting the parity checks of the SCo code in Table~\ref{Code1224}(b) to the right by two symbols and combining with the parity checks of the SCo code in Table~\ref{Code1224}(a) i.e., $q[i] = p^{\rm{I}}[i] + p^{\rm{II}}[i-2]$. Note that parity check symbols $p^{\rm{II}}[\cdot]$ do not interfere with the parity checks of user $1$ i.e., when $s[i]$ is erased, receiver 1 can recover $p^{\rm{I}}[i+1]$ and $p^{\rm{I}}[i+2]$ from $q[i+1]$ and $q[i+2]$ respectively by canceling $p^{\rm{II}}[\cdot]$ that combine with these symbols. It then recovers $s[i]$.  Likewise if $s[i]$ and $s[i-1]$ are erased, then receiver $2$ recovers $p^{\rm{II}}[i+1],\ldots, p^{\rm{II}}[i+4]$ from $q[i+3],\ldots,q[i+6]$ respectively by canceling out the interfering $p^{\rm{I}}[\cdot]$, thus yielding $T_2 = 6$. 

While the interference avoidance strategy illustrated above naturally generalizes to arbitrary values of $B$ and $T$, it is sub-optimal.  Table.~\ref{Code1225}(b) shows the DE-SCo construction that achieves the minimum possible delay of $T_2 = 5$. In this construction we first construct the parity checks $\cp^{\rm{II}}[i] = s_1[i-2]+ s_0[i-1]$ by combining the source sub-symbols along the opposite diagonal of the $(1,2)$ SCo code in Table~\ref{Code1224}(a). Note that $\check{x}(i) = (s[i], \cp^{\rm{II}}[i])$ is also a single user $(1,2)$ SCo code. We then shift the parity check stream to the right by $T+B = 3$ symbols and combine with $p^{\rm{I}}[i]$ i.e., $q[i] = p^{\rm{I}}[i] + \cp^{\rm{II}}[i-3]$. In the resulting code, receiver $1$ is still able to cancel the effect of $\cp^{\rm{II}}[\cdot]$ as before and achieve $T_1=2$. Furthermore at receiver $2$ if $s[i]$ and $s[i-1]$ are erased, then observe that receiver $2$ obtains $s_0[i]$ and $s_0[i-1]$ from $q[i+2]$ and $q[i+3]$ respectively and $s_1[i-1]$ and $s_1[i]$ from $q[i+4]$ and $q[i+5]$ respectively, thus yielding $T_2=5$ symbols. 

In the remainder of this paper we generalize the above construction to arbitrary values of $(B_i, T_i)$.

\section{Construction of DE-SCo}
In this section we describe the DE-SCo construction. We rely on several properties of the single user SCo explained in section~\ref{Background_Real}.
\begin{thm}
\label{thm:DE-SCo}
Let $(B_1,T_1)= (B,T)$ and suppose $B_2 = \al B$ where $\al$ is any integer that exceeds $1$. The minimum possible delay for any code of rate $R=\frac{T}{T+B}$ is
\begin{equation}
\label{eq:DE-SCoT2}T_2^\star = \al T + B,\end{equation}and is achieved by the optimal DE-SCo construction.
\end{thm}

\begin{figure}
		\centering
		\resizebox{\columnwidth}{!}{\includegraphics[trim = 0mm 110mm 0mm 10mm, clip]{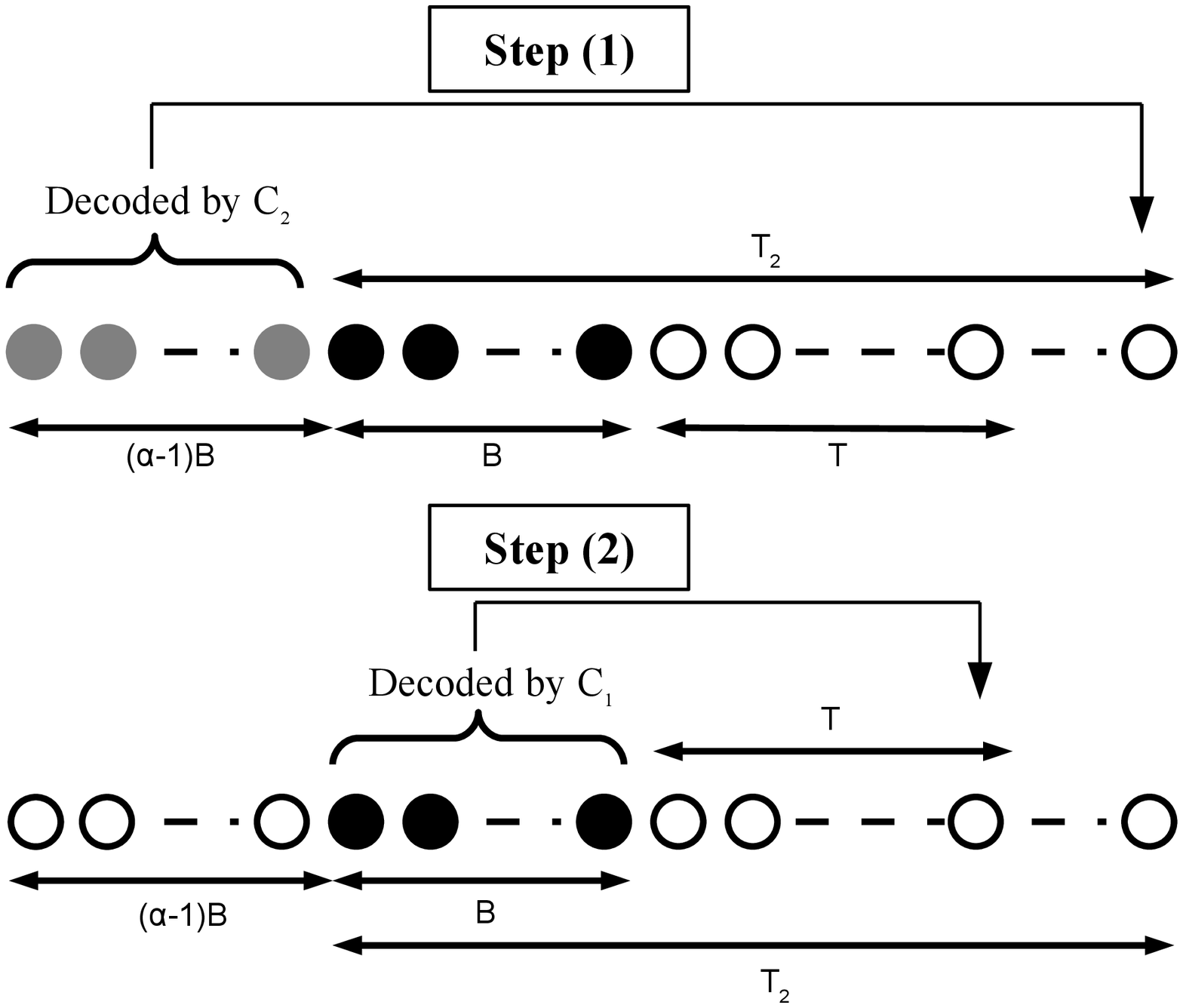}}
		\caption{One period illustration of the Periodic Erasure Channel for $T+B < T_2 \leq \al T+B$. White circles resemble unerased symbols. Black and Gray circles resemble erased symbols to be recovered using $\mathcal{C}_1$ and $\mathcal{C}_2$ respectively.
		\label{Periodic_Erasure_Channel_DESCo}}
\end{figure}

\vspace{-2em}
\begin{figure}
		\centering
		\resizebox{\columnwidth}{!}{\includegraphics[trim = 0mm 110mm 0mm 10mm, clip]{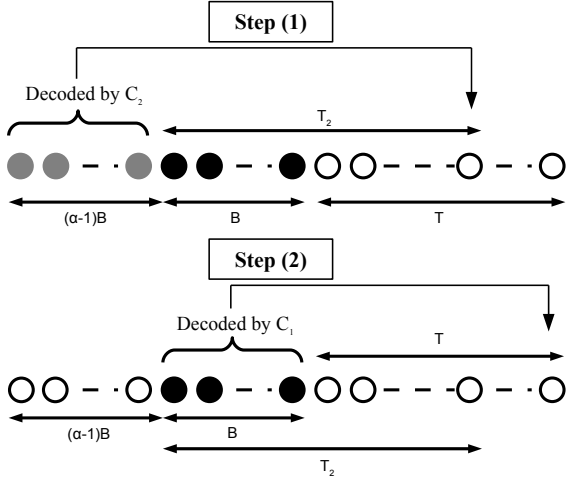}}
		\caption{One period illustration of the Periodic Erasure Channel for $T_2 \leq T+B$. White circles resemble unerased symbols. Black and Gray circles resemble erased symbols to be recovered using $\mathcal{C}_1$ and $\mathcal{C}_2$ respectively.
		\label{Periodic_Erasure_Channel_DESCo_2}}
\end{figure}

\subsection{Converse}
We first establish converse to theorem~\ref{thm:DE-SCo}. Consider any code that achieves $\{(B,T),(B_2,T_2)\}$  with $T_2 < T^\star_2$. The rate of this code is strictly less than $R= \frac{T}{T+B}$.

To establish this we separately consider the case when $T + B \le T_2 < \al T + B  $ and the case when $T_2 < T+B$. Let us assume the first case. 

As shown in Fig.~\ref{Periodic_Erasure_Channel_DESCo}, construct a periodic burst-erasure channel in which every period of $(\al -1)B + T_2$ symbols consists of a sequence of $\al B$ erasures followed by  a sequence of non-erased symbols.  Consider one period of the proposed periodic erasure channel with a burst erasure of length $\al B$ from time $t = 0,1,\dots,\al B - 1$ followed by a period of non-erasures for $t=\al B, \ldots,  \cT_{P_1} \triangleq (\al -1)B+T_2 -1$.  For time  $t = 0,\ldots, \cT_{P_1}$ the channel behaves identically to a burst-erasure channel with $\al B$ erasures. The first $(\al - 1)B$ erasures at time $t = 0,1,\dots,(\al-1)B-1$ can be recovered using decoder of user 2 with a delay of $T_2$ i.e., by time $\cT_{P_1}$ and hence the channel symbols $x[0],\dots,x[(\al-1)B-1]$ can also be recovered via~\eqref{Code_Function}.

It remains to show that the symbols at time $t = (\al-1)B,\dots,\al B -1$ are also recovered by time $\cT_{P_1}$. Note that since the channel symbols $x[0],\ldots, x[(\al -1)B-1]$ have been recovered, the resulting channel between times $t=0,\ldots, \al B-1$ is identical to a burst erasure channel with $B$ erasures between time $t=(\al-1)B,\ldots, \al B-1$.  The decoder of user 1 applied to this channel recovers the source symbols by time $\al B - 1 + T \le \cT_{P_1}$, which follows since $T_2 \ge T+B$. Thus all the erased channel symbols in the first period are recovered by time $\cT_{P_1}$. Since the channel introduces periodic bursts, the same argument can be repeated across all periods. Since the length of each period is $(\al -1)B + T_2$ and contains $\al B$ erasures,thus the capacity is upper bounded by $1-\frac{\al B}{(\al -1)B + T_2}$ which is less than $R = \frac{T}{T+B}$ if $T_2 < T_2^{\star}$.


For the other case with $T_2 < T + B$ shown in Fig.~\ref{Periodic_Erasure_Channel_DESCo_2}, the same argument applies except that the periodic channel has a period of $T+\alpha B$ symbols. Each period consists of a burst erasure of length $\al B$ from time $t = 0,1,\dots,\al B - 1$ followed by a period of non-erasures for $t=\al B, \ldots,  \cT_{P_2} \triangleq \al B+T-1$. The decoder of user 2 recovers the $(\al - 1)B$ erasures at time $t = 0,1,\dots,(\al-1)B-1$ with a delay of $T_2$ (i.e., by time $< \cT_{P_2}$) as $T_2 < T+B$. Furthermore, the decoder of user 1 recovers the $B$ erasures at time $t = (\al-1)B,\dots,\al B -1$ with a delay of $T$ symbols (i.e., by time $\al B + T -1 = \cT_{P_2}$). Now, the length of each period is $\al B + T$ with $T$ available symbols,  the rate is $\frac{T}{T + \al B}$ strictly smaller than $R$ as $\al > 1$ and the converse follows.

\subsection{Code Construction}
\label{subsec:DE-SCo_Construction}
For achievability of $T_2^\star$ in~\eqref{eq:DE-SCoT2} we construct the following code:\begin{itemize}
\item {\bf Construction of $\cC_1$}: Let $\cC_1$  be the single user $(B,T)$ SCo obtained by splitting each source symbol $s[i]$ into $T$ sub-symbols $(s_0[i],\ldots,s_{T-1}[i])$ and producing $B$ parity check sub-symbols $\bp^{\rm{I}} = (p^{\rm{I}}_0[i],\ldots, p^{\rm{I}}_{B-1}[i])$ at each time by combining the source sub-symbols along the main diagonal.

In other words, a $(T+B,T)$ LD-BEBC code is applied along the diagonal $\bvec^{\rm{I}}_{i} = (s_0[i],s_1[i],\dots,s_{T-1}[i+T-1])$ constructing the diagonal codeword $\dvec^{\rm{I}}_{i} = (s_0[i],\dots,s_{T-1}[i+T-1],p^{\rm{I}}_0[i+T],\dots,p^{\rm{I}}_{B-1}[i+T+B-1])$ where, from~\eqref{eq:ScoParity},
\begin{align}
p^{\rm{I}}_k[i]	&= \cA_k(s_0[i-T-k],\ldots, s_{T-1}[i-1-k]) \nonumber \\
								&= \cA_k(\bvec^{\rm{I}}_{i-T-k}) \nonumber \\
								&= s_k[i-T] + h_k(s_B[i-k-T+B],\ldots, \nonumber \\
								& s_{T-1}[i-k-1]),  \quad \quad \quad k=0,\ldots, B-1.
\end{align}

\item {\bf Construction of $\cC_2$}: Let $\cC_2$ be a  $((\al-1)B, (\al-1)T)$ SCo also obtained by splitting each source symbol into $T$ sub-symbols $(s_0[i],\ldots,s_{T-1}[i])$ and then constructing a total of $B$ parity checks $\bp^{\rm{II}}[i] = (p^{\rm{II}}_0[i],\ldots, p^{\rm{II}}_{B-1}[i])$ by combining the source sub-symbols along the opposite diagonal and with an interleaving step of size $\ell= (\al-1)$.

In other words, a $(T+B,T)$ LD-BEBC code is applied along the diagonal $\bvec^{\rm{II}}_{i} = (s_{T-1}[i-\ell(T-1)],s_{T-2}[i-\ell(T-2)],\dots,s_0[i])$ to construct a diagonal codeword $\dvec^{\rm{II}}_{i} = (s_{T-1}[i-\ell(T-1)],\dots,s_0[i],p^{\rm{II}}_0[i+\ell],\dots,p^{\rm{II}}_{B-1}[i+\ell B])$ where
\begin{align}
\label{eq:pB}
p^{\rm{II}}_k[i] 	&= \cB_k(s_0[i-\ell-k\ell],\ldots, s_{T-1}[i- \ell T - k\ell]) \nonumber \\
									&= \cB_k(\bvec^{\rm{II}}_{i-\ell-k\ell}) \nonumber \\
									&= s_{T-k-1}[i-\ell T] + h_k(s_{T-B-1}[i-\ell(k+T-B)], \nonumber \\  
									& \ldots, s_{0}[i-\ell(k+1)]), \quad k=0,\ldots, B-1,
\end{align}

\item{\bf Combination of Parity Checks of $\cC_1$ and $\cC_2$}: Introduce a shift $\Delta = T+B$ in the stream $p^{\rm{II}}[\cdot]$ and combine with the parity check stream $p^{\rm{I}}[\cdot]$ i.e.,  $\bq[i] = \bp^{\rm{I}}[i]+ \bp^{\rm{II}}[i-\Delta]$. The output symbol at time $i$ is $x[i] = (s[i],\bq[i])$ 
\end{itemize}

Throughout our discussion we refer to the non-urgent and urgent symbols of code $\cC_2$. The set of urgent symbols and non-urgent symbols are as stated in~\eqref{eq:UrgentNonUrgent_Defn_off}.  Also note that since there are $B$ parity check sub-symbols for every $T$ source sub-symbols it follows that the rate of the code is $\frac{T}{T+B}$.  

\begin{figure*}
		\centering
		\includegraphics[scale=0.75, trim = 0mm 0mm 0mm 0mm, clip]{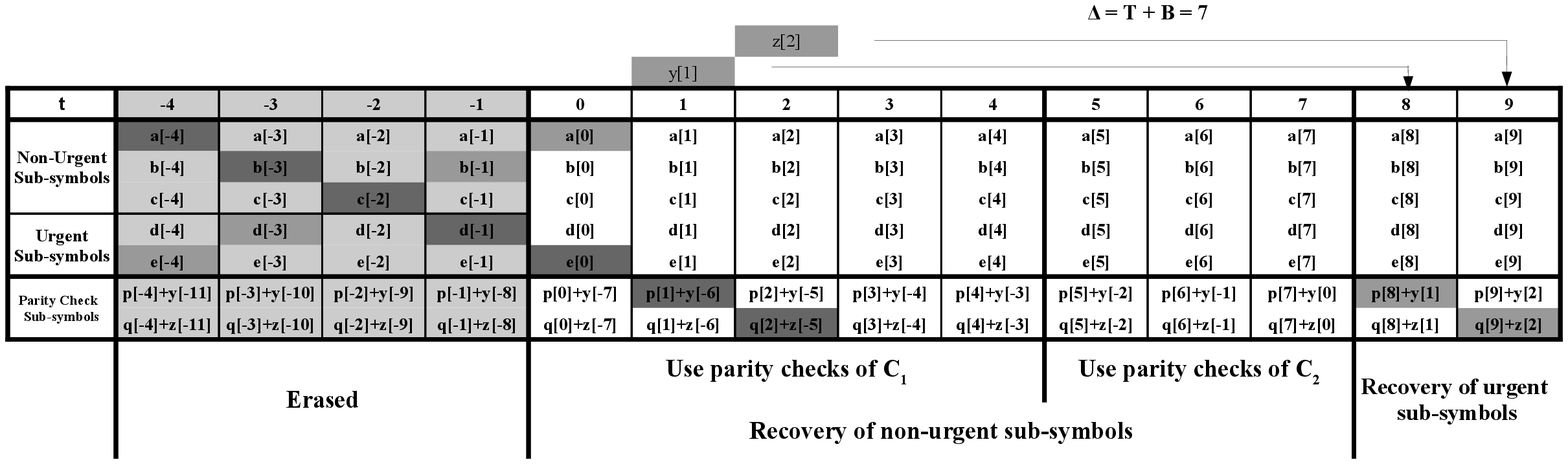}
		\caption{A $\{(2,5)-(4,12) \}$ DE-SCo code construction is given in the above figure. The parity check sub-symbols $p[t]$ and $q[t]$ of a $(2,5)$ SCo across the main diagonal is added to another $(2,5)$ SCo parity check sub-symbols $x[t]$ and $y[t]$ but applied across the opposite diagonal and shifted by $T+B = 7$ (i.e., the two parity check checks at time instant $t$ are $p[t]+x[t-7]$ and $q[t]+y[t-7]$).
		\label{DecodingExample}}
\end{figure*}

\subsection{Example}

Fig.~\ref{DecodingExample} illustrates the DE-SCo $\{(2,5), (4,12)\}$ construction. Each column represents one time-index between $[-4,9]$ shown in the top row of the table.  We assume that a burst-erasure occurs between time $[-4,-1]$ for user 2. Each source symbol is split into five sub-symbols $(a[.],b[.],c[.],d[.],e[.])$, each occupying one row. The first $T-B = 3$ of which, $a[.],b[.],c[.]$ are non-urgent sub-symbols and the rest $(d[.],e[.])$ are urgent. The next two rows denote the parity check sub-symbols.  The parity checks for $\cC_1$, generated by diagonal $\bd^{\rm{I}}_i$, (c.f.~\eqref{eq:ScoParity}) are marked $p[.]$ and $q[.]$ and are given by,
\begin{align}
\label{eq:parity_def_c1}
p[i] &= a[i-5] + c[i-3] + e[i-1] \nonumber \\
q[i] &= b[i-5] + d[i-3] + e[i-2].
\end{align}
The shaded two top rows show the parity checks $y[.]$ and $z[.]$, generated by the diagonal $\bd^{\rm{II}}_i$ for $\cC_2$ (c.f.~\eqref{eq:pB}),
\begin{align}
\label{eq:parity_def_c2}
y[i] &= e[i-5] + c[i-3] + a[i-1] \nonumber \\
z[i] &= d[i-5] + b[i-3] + a[i-2].
\end{align}
These parity checks are then shifted by $T+B=7$ slots and combined with the corresponding parity checks of $\cC_1$ as shown in Fig.~\ref{DecodingExample}.


We illustrate the decoding steps for user 2 as follows.
\begin{itemize}

\item[\textbf{(1)}] \textbf{Recover $\{ \bp^{\rm{II}}[t-\Delta] \}_{t \geq T}$:}\\
By construction of $\cC_1$ all the parity checks $\bp^{\rm{I}}[t]$ for $t\ge 5$ do not involve the erased sub-symbols. In particular the parity checks marked by $p[.]$ and $q[.]$ at $t \ge 5$  do not involve source sub-symbols before $t=0$ (c.f.~\eqref{eq:parity_def_c1}) and hence these can be canceled to recover the parity checks $y[.]$ and $z[.]$ for $t\ge 5$.

\item[\textbf{(2)}] \textbf{Upper-left triangle:}\\
The parity checks in step (1) enable us to recover the non-urgent erased sub-symbols in $\bd^{\rm{II}}_{-3}=(e[-7],d[-6],c[-5],b[-4],a[-3])$ and $\bd^{\rm{II}}_{-4} = (e[-8],d[-7],c[-6],b[-5],a[-4])$ which are $a[-4]$, $a[-3]$ and $b[-4]$ i.e., the upper-left triangle sub-symbols. We use the corresponding diagonal codewords, $\dvec^{\rm{II}}_{-3} = (e[-7],d[-6],c[-5],b[-4],a[-3],y[-2],z[-1])$ to recover $a[-3]$ and $b[-4]$ from the parity checks $y[-2]$ and $z[-1]$ and $\dvec_{-4} = (e[-8],d[-7],c[-6],b[-5],a[-4],y[-3],z[-2])$ to recover $a[-4]$ from the parity check $z[-2]$. We note that $a[-4]$ is recovered from $z[-2]$ at $t=5$ and not from $y[-3]$ which appears at $t=4$ and is not recovered in step (1). More generally, as we note later, the parity checks at $i+T$ and later suffice to recover symbols in this step.

\item[\textbf{(3)}] \textbf{Recover $\bp^{\rm{I}}[t]$ for $0 \leq t \leq T-1$:}\\
The sub-symbols recovered in step~(2)  suffice to recover all parity checks $\bp^{\rm{I}}[t]$ for $0 \le t \le 4$. Note that the relevant interfering parity checks from $\bp^{\rm{II}}[\cdot]$ in this interval is $y[-3] = e[-8] + c[-6] + a[-4]$. Since the only erased sub-symbol $a[-4]$ is already recovered in step~(2), these parity checks can be canceled. More generally as we show later, for the general case, our construction guarantees that the interfering parity checks $\bp^{\rm{II}}[\cdot]$ in the interval $0 \le t \le T-1$ only involve erased symbols from the upper left triangle, which are decoded in step (2).

\item[\textbf{(4)}] \textbf{Upper-right triangle:}\\
Since the diagonals $\bd^{\rm{I}}_{-2} = (a[-2],b[-1],c[0],d[1],e[2])$ and $\bd^{\rm{I}}_{-1} = (a[-1],b[0],c[1],d[2],e[3])$ involve two or fewer erasures, we can now recover these sub-symbols using parity checks of code $\mathcal{C}_1$ recovered in the previous step. In particular, the upper-right triangle source sub-symbols $a[-2]$, $b[-1]$ and $a[-1]$ can be recovered from $p[3]$, $q[4]$ and $p[4]$ respectively.

\item[\textbf{(5)}] \textbf{Recover non-urgent sub-symbols recursively:}\\
The remaining non-urgent sub-symbols need to be recovered in a recursive manner. Note that $\bd_{-3}^{\rm{I}} = (a[-3],b[-2],c[-1],d[0],e[1])$ has three erased sub-symbols. However, the first sub-symbol $a[-3]$ also belongs to $\bd^{\rm{II}}_{-3}$  and has already been recovered in step~(2). The remaining two sub-symbols, $b[-2]$ and $c[-1]$, can be recovered by the two available parity checks of code $\mathcal{C}_1$ in $\dvec_{-3}^{\rm{I}} = (a[-3],b[-2],c[-1],d[0],e[1],p[2],q[3])$, i.e., from $p[2]$ and $q[3]$. Similarly  $\bd^{\rm{II}}_{-2} = (e[-6],d[-5],c[-4],b[-3],a[-2])$ also has three erasures, but the upper-most sub-symbol $a[-2]$ also belongs to $\bd^{\rm{I}}_{-2}$ which has been recovered in step~(4). Hence the remaining erased sub-symbols in $\bd^{\rm{II}}_{-2}$, $c[-4]$ and $b[-3]$, can be recovered using the parity checks $y[-1]$ and $z[0]$ in $\dvec^{\rm{II}}_{-2} = (e[-6],d[-5],c[-4],b[-3],a[-2],y[-1],z[0])$.

At this stage it only remains to recover the two remaining non-urgent sub-symbols $c[-3]$ and $c[-4]$ by time $t=7$. These are recovered in the next step of the recursion.  Note that  the symbols $c[-2]$ and $d[-1]$ are the only remaining erased symbols on the diagonal  $\bd_{-5}^{\rm{I}}$ and are recovered from parity checks $p[1]$ and $q[2]$. Likewise, $c[-3]$ and $d[-4]$ are the only remaining erase symbols on the diagonal $\bd_{-1}^{\rm{II}}$ and can be recovered using the parity checks $y[0]$ and $z[1]$. Since $c[-3]$ is the non-urgent symbol, from Prop.~\ref{prop:UrgentNonUrgent_Decoding} it is recovered before $d[-4]$ using only $y[0]$. Thus both $c[-3]$ and $c[-4]$ are recovered by $t=7$.

\item[\textbf{(6)}] \textbf{Recover urgent sub-symbols:}\\
After recovering all non-urgent sub-symbols in the previous steps, we can directly recover the urgent ones (i.e., the bottom two rows) using parity checks $\bp^{\rm{II}}[t]$ for $ 8 < t \leq 11$.
\end{itemize}

We now study the general case.
\subsection{Decoding at User $1$}Suppose that the symbols at time $i-B,\ldots, i-1$ are erased by the channel of user $1$. User 1 first recovers parity checks $\bp^{\rm{I}}[i],\ldots, \bp^{\rm{I}}[i+T-1]$ from $\bq[i],\ldots, \bq[i+T-1]$ by canceling the parity checks $\bp^{\rm{II}}[\cdot]$ that combine with $\bp^{\rm{I}}[\cdot]$ in this period. Indeed at time $i+T-1$ the interfering parity check is $\bp^{\rm{II}}[i+T-\Delta-1]=\bp^{\rm{II}}[i-B-1]$, which clearly depends on the (non-erased) source sub-symbols before time $i-B$. All parity checks $\bp^{\rm{II}}[\cdot]$ before this time are also non-interfering. The erased source symbols can be recovered from $\bp^{\rm{I}}[i],\ldots, \bp^{\rm{I}}[i+T-1]$ by virtue of code $\cC_1$.

\subsection{Decoding at User $2$}
\label{subsubsec:user2Dec}

Suppose that the symbols at times $i-\al B,\ldots, i-1$ are erased for receiver 2.  Let $\cT \stackrel{\Delta}{=} i-\al B + T^\star_2$. We use parity checks at time $i \le t \le \cT-1$ to recover $\{\bs^N[\tau]\}_{\tau=i-\al B}^{i-1}$ in the first five steps where $ \bs^N[\tau]=(s_0[\tau],\ldots, s_{T-B-1}[\tau])$ denote the set of \emph{non-urgent} sub-symbols for $\cC_2$. In the last step, we use parity checks at time $t \geq \cT$ to recover the set of \emph{non-urgent} sub-symbols for $\cC_2$, $ \bs^U[\tau]=(s_{T-B}[\tau],\ldots, s_{T-1}[\tau])$.


\begin{itemize}
\item[\textbf{(1)}] \textbf{Recover $\{ \bp^{\rm{II}}[t-\Delta] \}_{t \geq i+T}$:}\\
For $t \ge i+T$, the decoder recovers parity check $\bp^{\rm{II}}[t-\Delta]$ from $\bq[t]$ by canceling the parity checks $\bp^{\rm{I}}[t]$ which depend only on (non-erased) source symbols at time $i$ or later as via~\eqref{eq:ScoParity} the memory in $\cC_1$ is limited to previous $T$ symbols.  Consequently the parity check symbols $\{\bp^{\rm{I}}[t]\}_{t \geq i+T}$ depend only on source sub-symbols after time $i$. Hence these parity checks can be canceled.

\item[\textbf{(2)}] \textbf{Upper-left triangle:}\\
In this step, the decoder recovers the non-urgent sub-symbols in $\bd^{\rm{II}}_{i-\al B},\ldots, \bd^{\rm{II}}_{i-B-1}$ using the parity check symbols $\{\bp^{\rm{II}}[t-\Delta]\}_{t=i+T}^{\CapTau-1}$.
Clearly these vectors are affected by at most $(\al - 1)B$ erasures between times $i-\al B,\ldots, i-B-1$.  Furthermore, the corresponding parity checks $\{\bp^{\rm{II}}[t-\Delta]\}_{t\ge i+T}\equiv \{ \bp^{\rm{II}}[t] \}_{t \geq i-B}$ have been recovered in step (1). By construction $\cC_2$ can recover the erased source sub-symbols in the stated diagonal vectors. Furthermore by applying Prop.~\ref{prop:UrgentNonUrgent_Decoding}, the non-urgent sub-symbols are recovered from the first $(\al-1)(T-B)$ parity check columns. Taking into account the shift of $\Delta = T+B$,  it follows that all the non-urgent source sub-symbols are recovered by time $i+T+(\al-1)(T-B)-1 = \CapTau -1$.

\item[\textbf{(3)}] \textbf{Recover $\bp^{\rm{I}}[t]$ for $i \leq t \leq i+T-1$:}\\
We consider the last column of parity checks, $\bq[i+T-1] = \bp^{\rm{I}}[i+T-1] + \bp^{\rm{II}}[i-B-1]$. From \eqref{eq:pB}, for $k = 0,1,\dots,B-1$ we have,
\small
\begin{align*}
p^{\rm{II}}_k[i-B-1] 	&= \cB_k(\bd^{\rm{II}}_{i-B-(\al-1)(k+1)-1}) \nonumber \\
											&= s_{T-k-1}[i-B-1-(\al-1)T] \\ & + h_k(s_{T-B-1}[i-B-1 - (\al-1)(T-B+k)],\\ & \ldots, s_{0}[i-B-1-(k+1)(\al-1)]).
\end{align*}
\normalsize
 Thus the only urgent sub-symbols involved in $\bp^{\rm{II}}[i-B-1]$ are at time  $t= i-B-1-(\al-1)T$,  which are unerased. Moreover, the non-urgent sub-symbols involved are those of $\bd^{\rm{II}}_{i-B-(\al-1)(k+1)-1}$ which have already been recovered in step (2). Thus, it follows that we can reconstruct $\bp^{\rm{II}}[i-B-1]$. A similar argument can be used to show that we can recover all the columns $\bp^{\rm{II}}[i-B-T],\ldots,\bp^{\rm{II}}[i-B-1]$, cancel their effect on $\bq[i],\ldots,\bq[i+T-1]$ and recover $\bp^{\rm{I}}[i],\ldots,\bp^{\rm{I}}[i+T-1]$.


\item[\textbf{(4)}] \textbf{Upper-right triangle:}\\
In this step, the decoder recovers the non-urgent sub-symbols in $\bd^{\rm{I}}_{i-1},\ldots, \bd^{\rm{I}}_{i-B}$ using the parity checks $\bp^{\rm{I}}[i],\ldots, \bp^{\rm{I}}[i+T-1]$.
Step~(4) follows in a similar way to step~(2). The  diagonal vectors $\bd^{\rm{I}}_{i-B},\dots,\bd^{\rm{I}}_{i-1}$ spanning the upper-right triangle of the erased source sub-symbols are affected  by a burst erasure of length $B$ between times $i-B,\dots,i-1$.  Furthermore, the corresponding parity checks $\{ \bp^{\rm{I}}[t] \}_{i \leq t < i+T}$  recovered earlier are capable of recovering the erased source sub-symbols in these  diagonal vectors by at most time $i+T-1 < \CapTau$.

\item[\textbf{(5)}] \textbf{Recover non-urgent sub-symbols recursively:}\\
For each $k \in \{1,\ldots, T-B-1\}$ recursively recover the remaining non-urgent sub-symbols as follows:
\begin{itemize}
\item[(Ind. 1)] Recover the non-urgent sub-symbols in $\bd^{\rm{I}}_{i-B-k}$ using the non-urgent sub-symbols in $\{\bd^{\rm{II}}_j\}_{j \le i + (k-1)(\al-1)-B-1}$ and parity checks $\bp^{\rm{I}}[\cdot]$ between $i\le t< i+T$.
\item[(Ind. 2)] Recover the non-urgent sub-symbols in $\bd^{\rm{II}}_{i-B+(k-1)(\al-1)},\ldots, \bd^{\rm{II}}_{i-B+k(\al-1)-1}$ using $\{\bd^{\rm{I}}_j\}_{j\ge i-B-(k-1)}$ and the parity checks  $\bp^{\rm{II}}[\cdot]$ between $i+T \le t < \CapTau$. 
\end{itemize}
Once this recursion terminates, all the non-urgent sub-symbols $\{\bs^N[\tau]\}_{\tau=i-\al B}^{i-1}$ are recovered by time $\CapTau-1$.

We establish the claim of the recursion using induction. Consider the case when $k=1$. According to Ind.~$1$ the non-urgent sub-symbols $\{\bd^{\rm{II}}_j\}_{j\le i-B-1}$ are available (from step 1). To recover $\bd^{\rm{I}}_{i-B-1}$, note that the only erased sub-symbol in this vector before time $i-B$ is $s_0[i-B-1]$ which has already been recovered in $\bd^{\rm{II}}_{i-B-1}$. Hence the parity checks of $\cC_1$ at the times $i,\dots,i+T-1$ suffice to recover the remaining sub-symbols.  According to Ind.~2 the non-urgent sub-symbols in $\{\bd_j^{\rm{I}}\}_{j\ge i-B}$ have been recovered in step~(4). Furthermore in vectors $\bd^{\rm{II}}_{i-B},\ldots, \bd^{\rm{II}}_{i-B+\al-2}$ the only erased sub-symbols after time $i-B-1$ are $s_0[i-B],\ldots, s_0[i-B+\al-2]$, which are available from $\{\bd_j^{\rm{I}}\}_{j\ge i-B}$.  Thus the parity checks $\bp^{\rm{II}}[\cdot]$ can be used to recover the remaining non-urgent sub-symbols in these vectors. 

Next suppose the statement holds for some $t =k$. We establish that the statement holds for $ t= k+1$.  In Ind.~$1$ the vector of interest is,
\begin{align*}
\bd^{\rm{I}}_{i-B-(k+1)} = &(s_0[i-B-(k+1)],...,s_{k}[i-B-1],..., \\
& s_{T-1}[i-B-k+(T-2)]).
\end{align*}
The erased elements in the interval $i-\al B,\dots,i-B-1$ are $s_j[i-B-k+j-1]$ for $j=0,\dots,k$. Note that $s_j[i-B-k+j-1]$ is precisely the $j-$th sub-symbol in the diagonal vector $\bd^{\rm{II}}_{i-B-k+\al j-1}$.  Furthermore the diagonals of interest $\bd^{\rm{II}}_{i-B-k-1},\dots,\bd^{\rm{II}}_{i-B+(\al-1)k-1}$, already visited in Ind.~$2$ in the $k$-th recursion.   Hence the remaining sub-symbols are recovered using the parity checks of $\cC_1$. 

For Ind.~$2$, the first vector of interest at step $k+1$ is
\begin{align*}
& \bd^{\rm{II}}_{i-B+ k(\al-1)} = (s_0[i-B+k(\al-1)],..., \\
& \quad \quad \quad s_{k}[i-B],s_{k+1}[i-B-(\al-1)],...).
\end{align*}Note that the  sub-symbols  $s_0[.],\ldots, s_k[.]$ above, also belong to vectors $\bd^{\rm{I}}_{i-B+(\al-1)k},\ldots, \bd^{\rm{I}}_{i-B-k},$ and are recovered in Ind.~$1$ by the $k-$th step.  Since the remaining erased symbols span the interval $[i-\al B, i-B)$  the parity checks $\{\bp^{\rm{II}}[\cdot]\}_{t\ge i-B }$ recovered in step~(3)  can be used to recover these erased symbols.

Likewise, the last vector of interest at step $k+1$ is
\begin{align*}
& \bd^{\rm{II}}_{i-B+ (k+1)(\al-1)-1} = (s_0[i-B+(k+1)(\al-1)-1], \\ 
& \quad \quad \ldots, s_{k}[i-B+(\al-1)-1],s_{k+1}[i-B-1],...).
\end{align*}
Note that the  sub-symbols  $s_0[.],\ldots, s_k[.]$ above, also belong to vectors  $\bd^{\rm{I}}_{i-B+(k+1)(\al-1)-1},$ $\ldots$  $\bd^{\rm{I}}_{i-B+(\al-1)-k-1}$ which   are recovered in Ind.~$1$ by step number $k+1-(\al-1) < k+1$. Since the remaining erased symbols span the interval $[i-\al B, i-B)$  the parity checks $\{\bp^{\rm{II}}[\cdot]\}_{t\ge i-B }$ recovered in step~(3)  can be used to recover these erased symbols. 

It only remains to show that the non-urgent symbols in the diagonal $\bd^{\rm{II}}$ are all recovered before time $\CapTau$. From Proposition.~\ref{prop:UrgentNonUrgent_Decoding} all the non-urgent sub-symbols are recovered using the first $(\al-1)(T-B)$ columns of the parity checks $\{\bp^{\rm{II}}[\cdot]\}_{t\ge i-B }$.  Since these parity checks are shifted by $T+B$, the fall in the interval $i+T,\dots,i+T+(\al-1)(T-B)-1 = \CapTau-1$. Thus only the parity checks before time $\CapTau$ are required to recover the non-urgent source sub-symbols. \\

This completes the claim in the Ind.~1 and Ind.~2.  We finally show that all the non-urgent erased source sub-symbols are recovered at $k = T-B-1$. Because of the recovery along the diagonals, it suffices to show that the lower left most non-urgent sub-symbol in the region $i-B,\dots,i-1$ i.e., $s_{T-B-1}[i-B]$ is an element of $\bd^{\rm{I}}_{i-B-k} = \bd^{\rm{I}}_{i-T+1}$ which is clear from the definition of $\bd^{\rm{I}}_{i}$ at $i-T+1$ as,
\small
\begin{align*}
\bd^{\rm{I}}_{i-T+1} = (s_0[i-T+1],\dots,s_{T-B-1}[i-B],\dots,s_{T-1}[i]).
\end{align*}
\normalsize
Similarly, we need to show that $\bd^{\rm{II}}_{i-B+k(\al-1)-1} = \bd^{\rm{II}}_{i-B+(T-B-1)(\al-1)-1}$ contains the lower right most non-urgent sub-symbol in the region $i-\al B,\dots,i-B-1$ i.e., $s_{T-B-1}[i-B-1]$. This too immediately follows by applying the definition of $\bd^{\rm{II}}_i$ at time $i-B+(T-B-1)(\al-1)-1$ as,
\footnotesize
\begin{align*}
\bd^{\rm{II}}&_{i-B+(T-B-1)(\al-1)-1} = (s_0[i-B+(T-B-1)(\al-1)-1], \\ &\dots,s_{T-B-1}[i-B-1],\dots,s_{T-1}[i-\al B - 1]).
\end{align*}
\normalsize

\item[\textbf{(6)}] \textbf{Recover urgent sub-symbols:}\\
Finally, the decoder recovers urgent sub-symbols $\bs^U[\tau] = (s_{T-B}[\tau],\ldots, s_{T-1}[\tau])$ for $i-\al B \le \tau < i$ at time $t = \tau + T^\star_2$ using the parity check symbols $\bp^{\rm{II}}[t]$ and the previously decoded non-urgent sub-symbols. 
We establish this claim as follows. After recovering all the non-urgent source sub-symbols $\{\bs^N[\tau]\}_{\tau=i-\al B}^{i-1}$, 	we can directly apply the construction of $\cC_2$ to recover the urgent sub-symbols $\{\bs^U[\tau]\}_{\tau=i-\al B}^{i-1}$ using parity checks $\bp^{\rm{II}}[\cdot]$ within a delay of $T_2^{\star}$.\\
\end{itemize}

{\bf Note on Computational Complexity}: We note that a DE-SCo encoder and decoder are of a polynomial complexity as the DE-SCo constructions are built upon a linear convolutional code with finite memory. Specifically, going through the  steps (1)-(4) of the DE-SCo decoder, we can conclude that since every erased sub-symbol is processed at-most once, the complexity of any step is no more than  $\alpha B T$. In step (5) we use a  recursive decoder that terminates in $T+B-1$ recursions. Also each step has at-most $\alpha B T$ and thus the complexity is polynomial in $\al$, $B$ and $T$.

\section{General Values of $\al$}

In this section, we show that DE-SCo codes $\{(B,T), (\al B,\al T + B)\}$ can be constructed for any non-integer value of $\alpha$ such that $B_2 =\alpha B$ is an integer. For any $\alpha = \frac{B_2}{B} > 1$, let $\alpha = \frac{a}{b}$ where $a$ and $b$ are integers and $\frac{a}{b}$ is in the simplest form.

\subsection{DE-SCo Construction}

We introduce suitable modifications to the construction given in the previous section.  Clearly since $\frac{a}{b}$ is in simplest form $B$ must be an integer multiple of $b$ i.e., $B_0 = \frac{B}{b} \in \mathbb{N}$. We first consider the case when $T$ is also an integer multiple of $b$ i.e., $T_0 = \frac{T}{b} \in \mathbb{N}$. The case when $T$ is not an integer multiple, can be dealt with by a suitable source expansion, as outlined at the end of the section.  
\begin{itemize}
\item Let $\cC_1$  be the single user $(B,T) = (b B_0, b T_0)$ SCo obtained by splitting each source symbol $s[i]$ into $T_0$ sub-symbols $(s_0[i],\ldots,s_{T_0 -1}[i])$ and producing $B_0$ parity check sub-symbols $\bp^{\rm{I}} = (p^{\rm{I}}_0[i],\ldots, p^{\rm{I}}_{B_0 -1}[i])$ at each time by combining the source sub-symbols along the main diagonal with an interleaving step of size $b$ i.e.,\begin{equation}p^{\rm{I}}_k[i] \!=\! \cA_{k}(s_0[i-bT_0 -kb],\ldots, s_{T_0 -1}[i-b-kb])\end{equation}

\item  Let $\cC_2$ be a  $((\al-1)B, (\al-1)T) = ((a-b)B_0, (a-b)T_0)$ SCo also obtained by splitting the source symbols into $T_0$ sub-symbols $(s_0[i],\ldots,s_{T_0 -1}[i])$ and then constructing a total of $B_0$ parity checks $\bp^{\rm{II}} = (p^{\rm{II}}_0[i],\ldots, p^{\rm{II}}_{B_0 -1}[i])$ by combining the source sub-symbols along the opposite diagonal and with an interleaving step of size $\ell = (a-b)$ i.e.,\begin{equation}p^{\rm{II}}_k[i] = \cB_k(s_0[i-\ell-k\ell],\ldots, s_{T_0 -1}[i- \ell {T_0} - k\ell]).\end{equation}

\item Introduce a shift $\Delta = T+B = b(T_0 + B_0)$ in the stream $p^{\rm{II}}[\cdot]$ and combine with the parity check stream $p^{\rm{I}}[\cdot]$ i.e.,  $\bq[i] = \bp^{\rm{I}}[i]+ \bp^{\rm{II}}[i-\Delta]$. The output symbol at time $i$ is $x[i] = (s[i],\bq[i])$. 
\end{itemize}

\subsection{Decoding}
The decoding steps is analogous to the case when $\al$ is integer. We sketch the main steps. As before the decoding is done along the diagonal vectors $\bd^{\rm{I}}_i \!=\!(s_0[i],\ldots, s_{T_0 -1}[i+(T_0 -1)b])$, $\bd^{\rm{II}}_i \!= \!(s_0[i],\ldots,s_{T_0 -1}[i\! -( {T_0}\! -1)\ell\!]).$

\subsubsection*{Decoding at User $1$}For the first user, the same argument applies  as in previous section i.e., a shift of $\Delta = b(T_0 + B_0)$ in $\bp^{\rm{II}}[\cdot]$ guarantees that user 1 can cancel the interfering parity checks to recover the $\bp^{\rm{I}}[\cdot]$ stream of interest. 

\subsubsection*{Decoding at User $2$}
We verify that steps in section~\ref{subsubsec:user2Dec} continue to apply.   A little examination shows that the claims~(1)---(4) as well as the proofs in the previous case follow immediately as they hold for an arbitrary interleaving step for $\cC_2$ and do not rely on the interleaving step of $\cC_1$ being $1$. The induction step needs to be modified to reflect that the interleaving step size of $\cC_1$ is $b > 1$.

For each $k \in \{1,\ldots, T-B-1\}$ recursively recover the remaining non-urgent sub-symbols as follows:
\begin{itemize}
\item {\bf Ind. 1} Recover the non-urgent sub-symbols in $\bd^{\rm{I}}_{i-B-(k-1)b-1},\dots,\bd^{\rm{I}}_{i-B-kb}$ using the non-urgent sub-symbols in $\{\bd^{\rm{II}}_j\}_{j \le i + (k-1)(a-b)-B-1}$ and parity checks $\bp^{\rm{I}}[\cdot]$ between $i\le t< i+T$.
\item {\bf Ind. 2} Recover the non-urgent sub-symbols in $\bd^{\rm{II}}_{i-B+(k-1)(a-b)},\ldots, \bd^{\rm{II}}_{i-B+k(a-b)-1}$ using $\{\bd^{\rm{I}}_j\}_{j\ge i-B-(k-1)b}$ and the parity checks  $\bp^{\rm{II}}[\cdot]$ between $i+T \le t < \CapTau$. 
\end{itemize}
Once this recursion terminates, all the non-urgent sub-symbols $\{\bs^N[\tau]\}_{\tau=i-\al B}^{i-1}$ are recovered by time $\CapTau-1$. The proof of this recursion is also similar to the previous section and will be omitted.

Finally the assumption that $T$ is a multiple of $b$ (i.e. $\alpha T$ is an integer) can be relaxed through a source pseudo-expansion approach as follows:
\begin{itemize}
\item Split each source symbol into $n T$ sub-symbols $s_0[i],\dots,s_{n T -1}[i]$  where $n$ is the smallest integer such that $n \alpha T$ is an integer.
\item Construct an expanded source sequence $\tilde{s}[.]$ such that $\tilde{s}[ni+r] = (s_{r T}[i],\dots,s_{(r+1)T - 1}[i])$ where $r \in \{ 0,\dots,n-1\}$.
\item We apply a DESCo code with parameters $\{ (n B,n T)-(n \alpha B, n (\alpha T + B)) \}$ to $\tilde{s}[.]$ using the earlier construction.
\end{itemize}
Notice that since the channel  introduces a total of $B_i$ erasures on the original input there will be $nB_i$ erasures on the expanded stream. These will be decoded with a delay of $nT_i$ on the expanded stream, which can be easily verified to incur a delay of $T_1$ and $\lceil T_2 \rceil$ on the original stream for user 1 and 2 respectively.

\section{Numerical Results}

\begin{figure*}
\begin{minipage}[b]{0.5\linewidth}
\centering
\begin{tabular}{c}
\resizebox{\columnwidth}{!}{\includegraphics[scale=0.9]{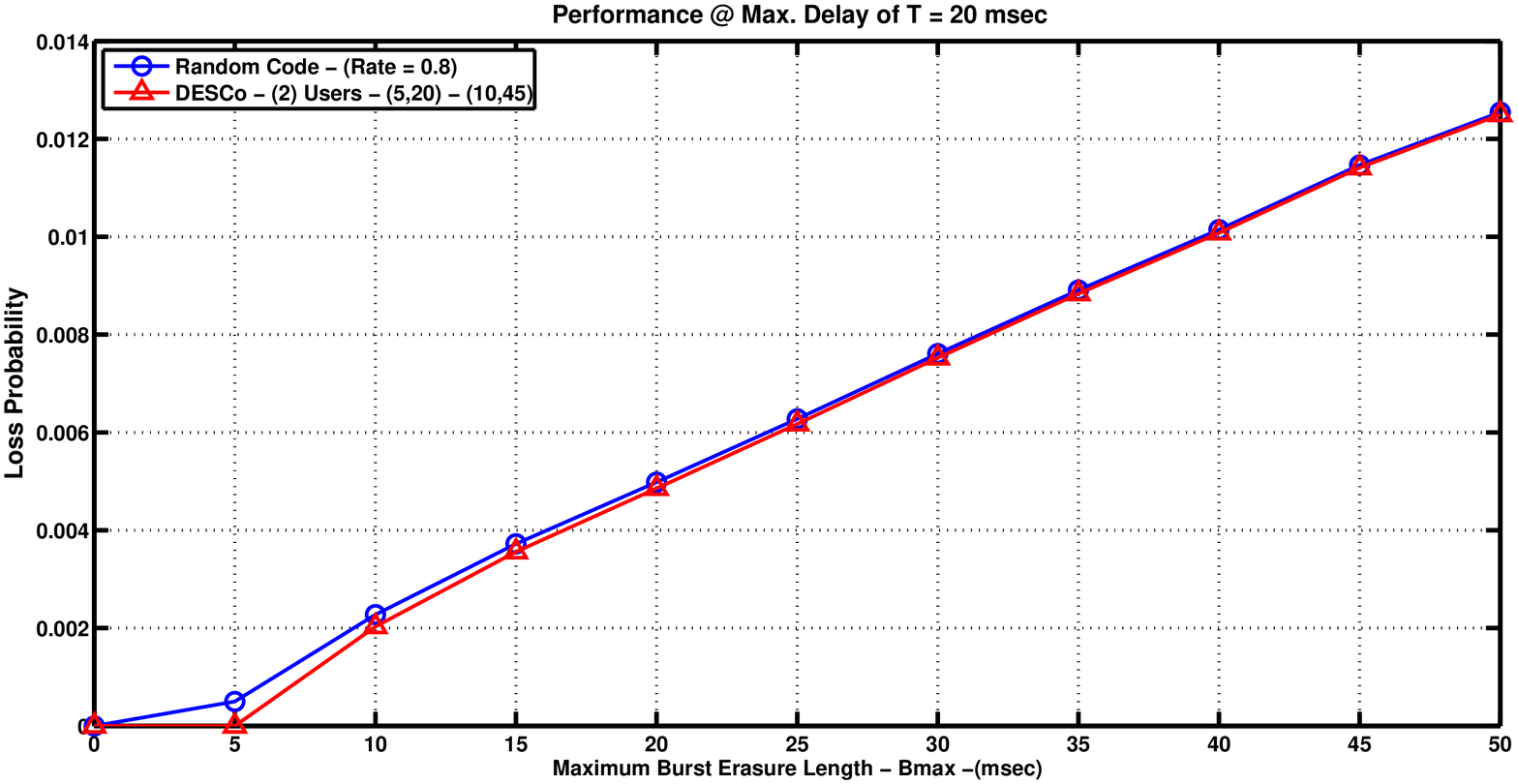}} \\
\resizebox{\columnwidth}{!}{\includegraphics[scale=0.9]{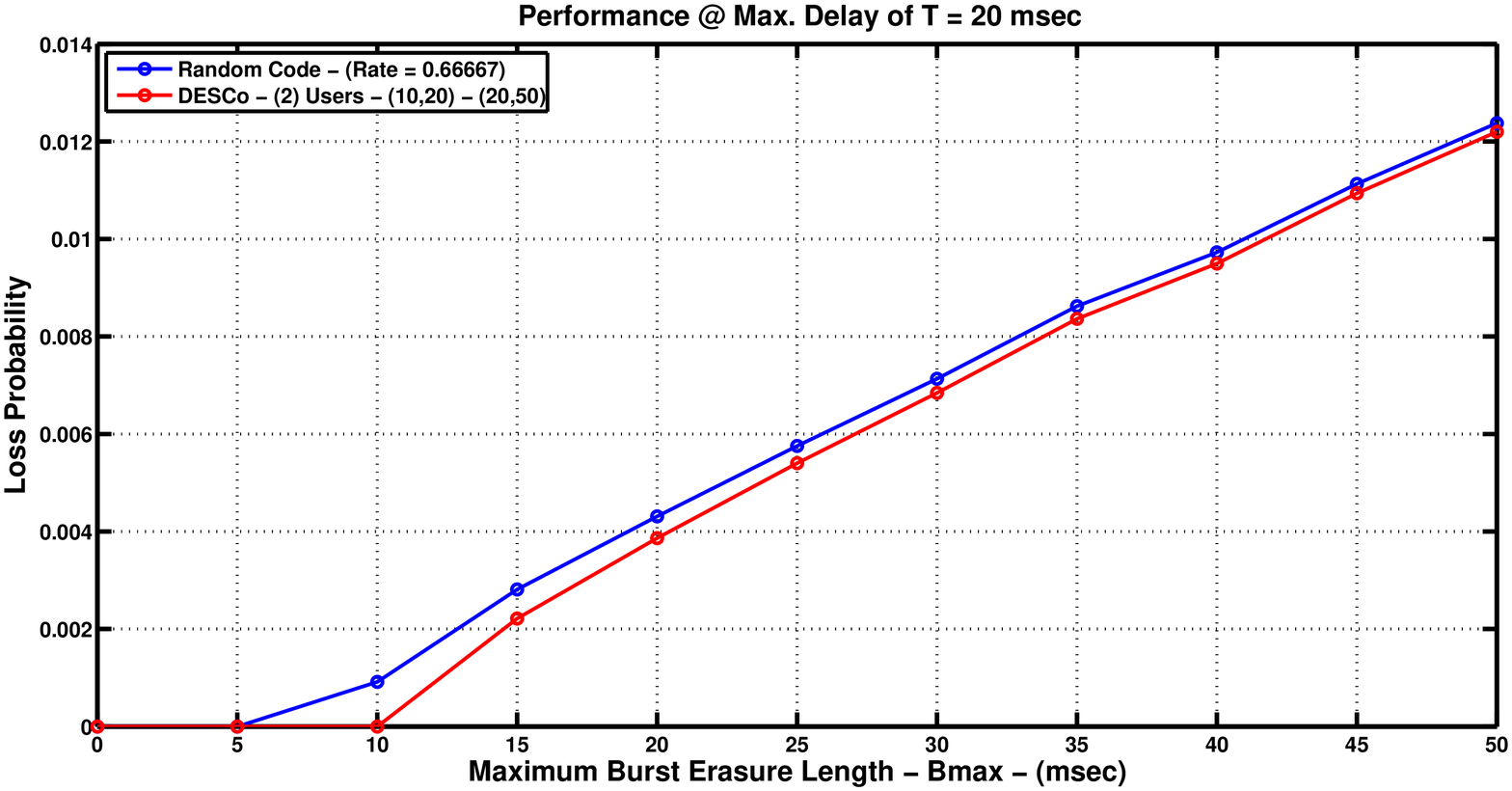}} \\
\resizebox{\columnwidth}{!}{\includegraphics[scale=0.9]{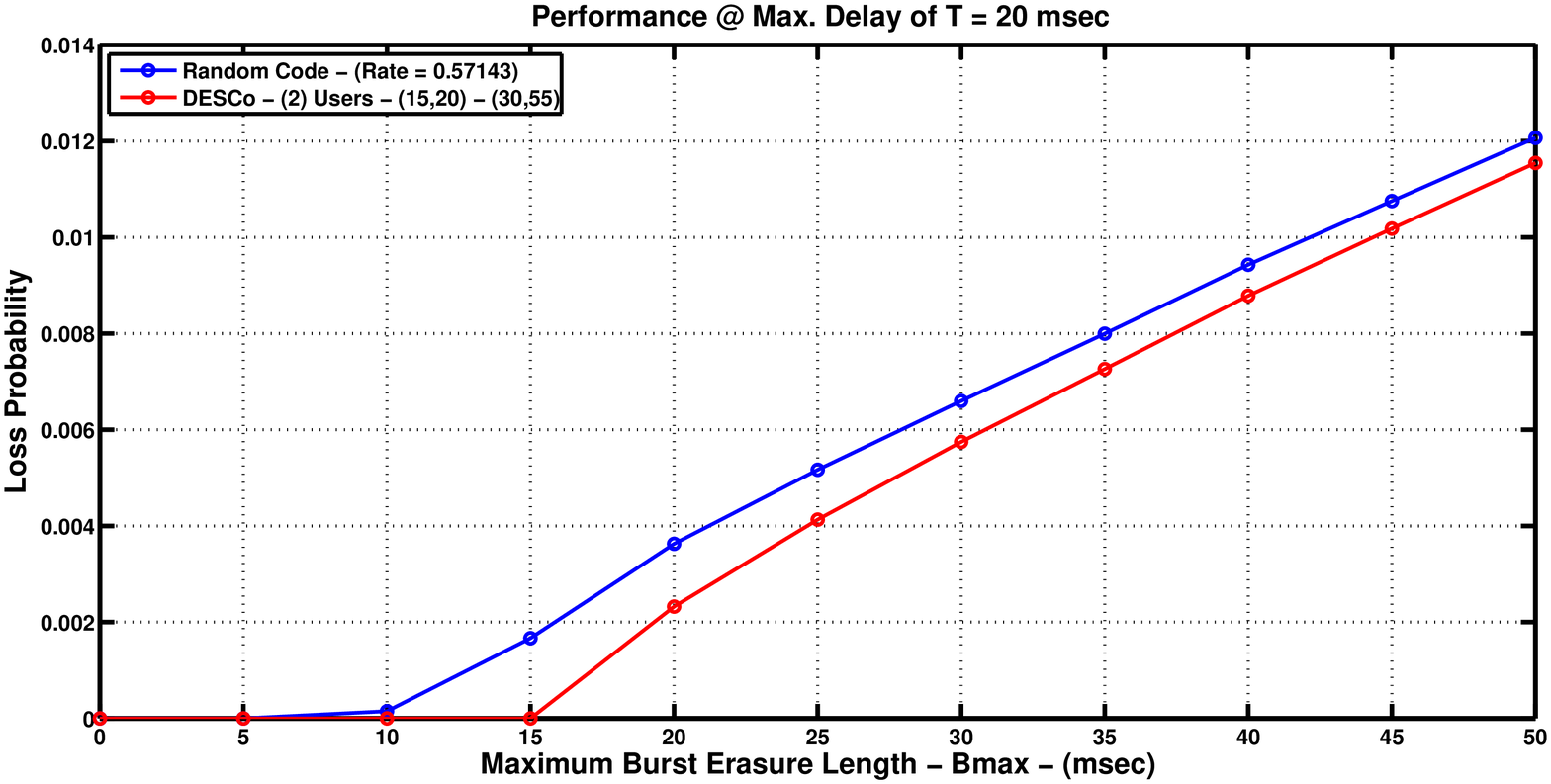}} \\
\resizebox{\columnwidth}{!}{\includegraphics[scale=0.9]{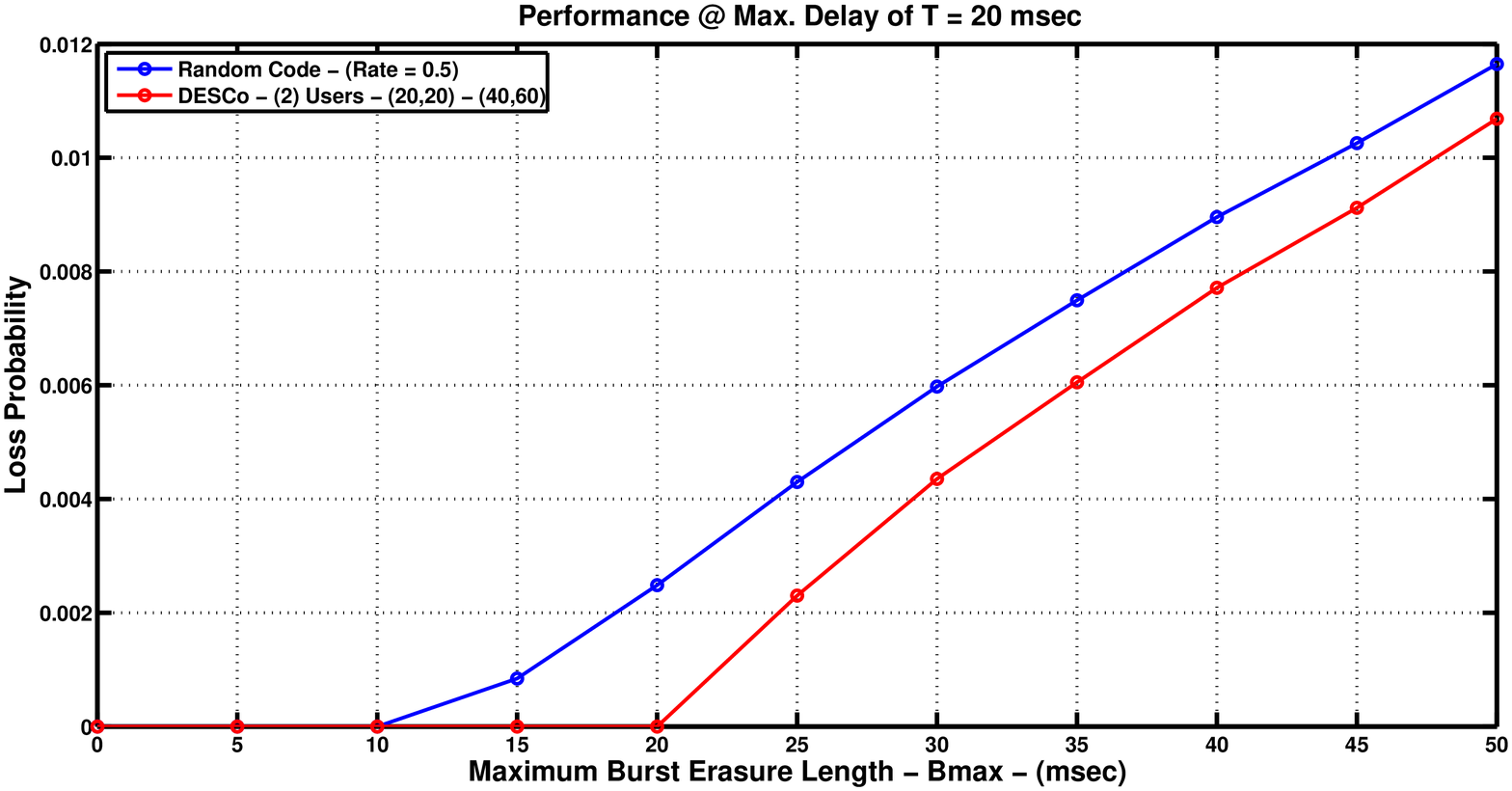}}
\end{tabular}
\caption{Loss Probability at the first receiver.}
\label{NR_User1}
\end{minipage}
\hspace{0.5cm}
\begin{minipage}[b]{0.5\linewidth}
\centering
\begin{tabular}{c}
\resizebox{\columnwidth}{!}{\includegraphics[scale=1]{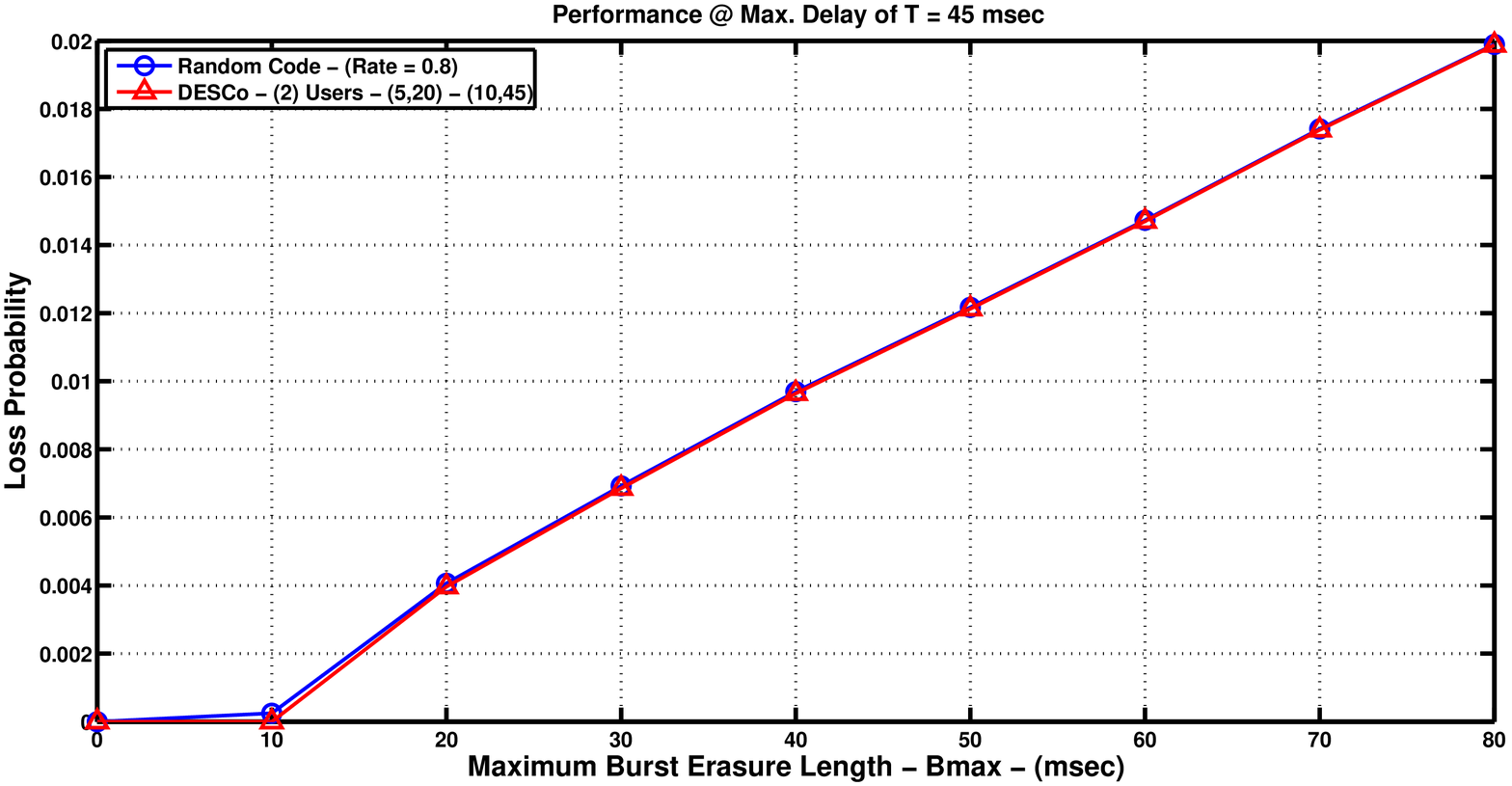}} \\
\resizebox{\columnwidth}{!}{\includegraphics[scale=1]{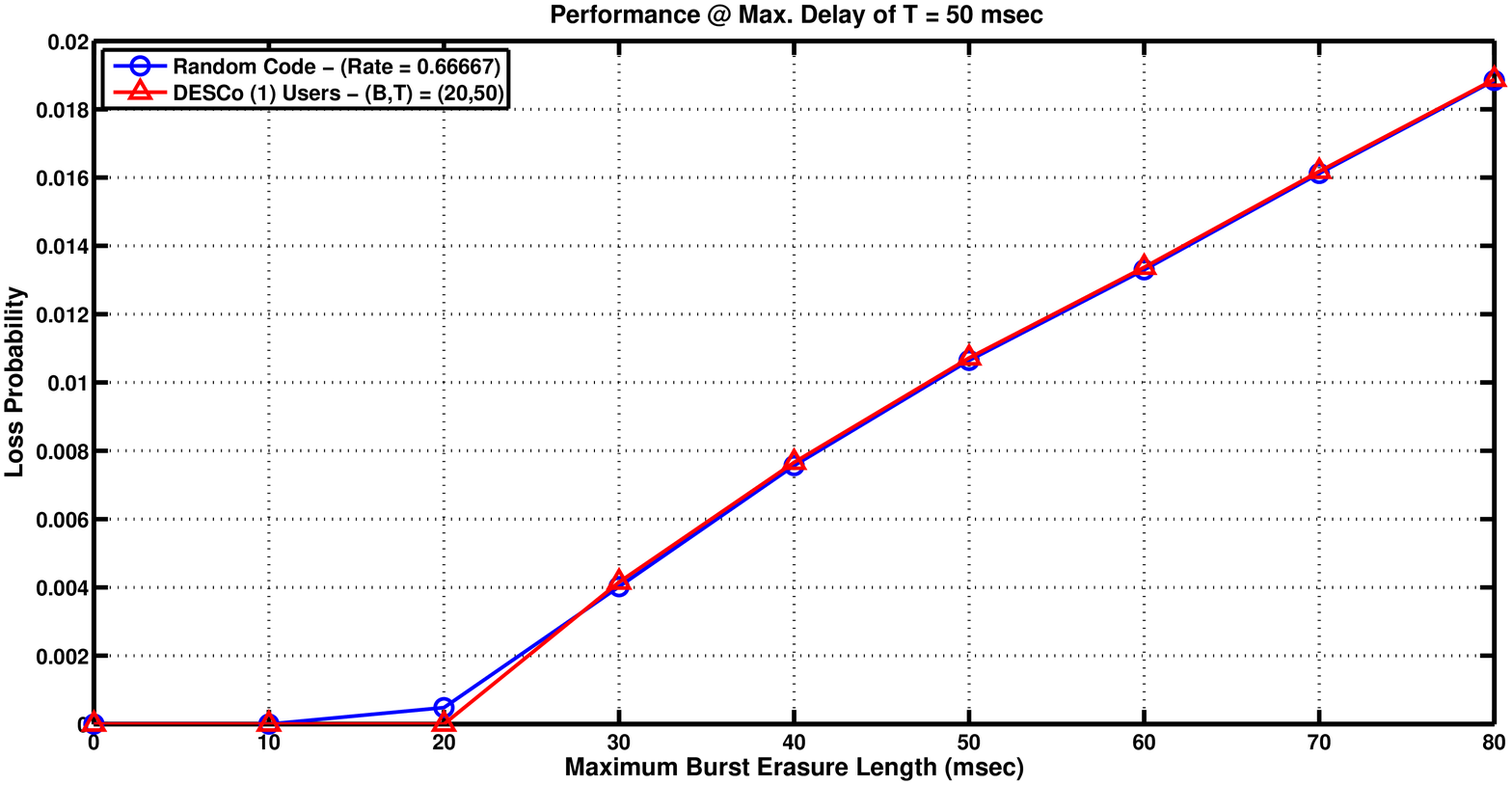}} \\
\resizebox{\columnwidth}{!}{\includegraphics[scale=1]{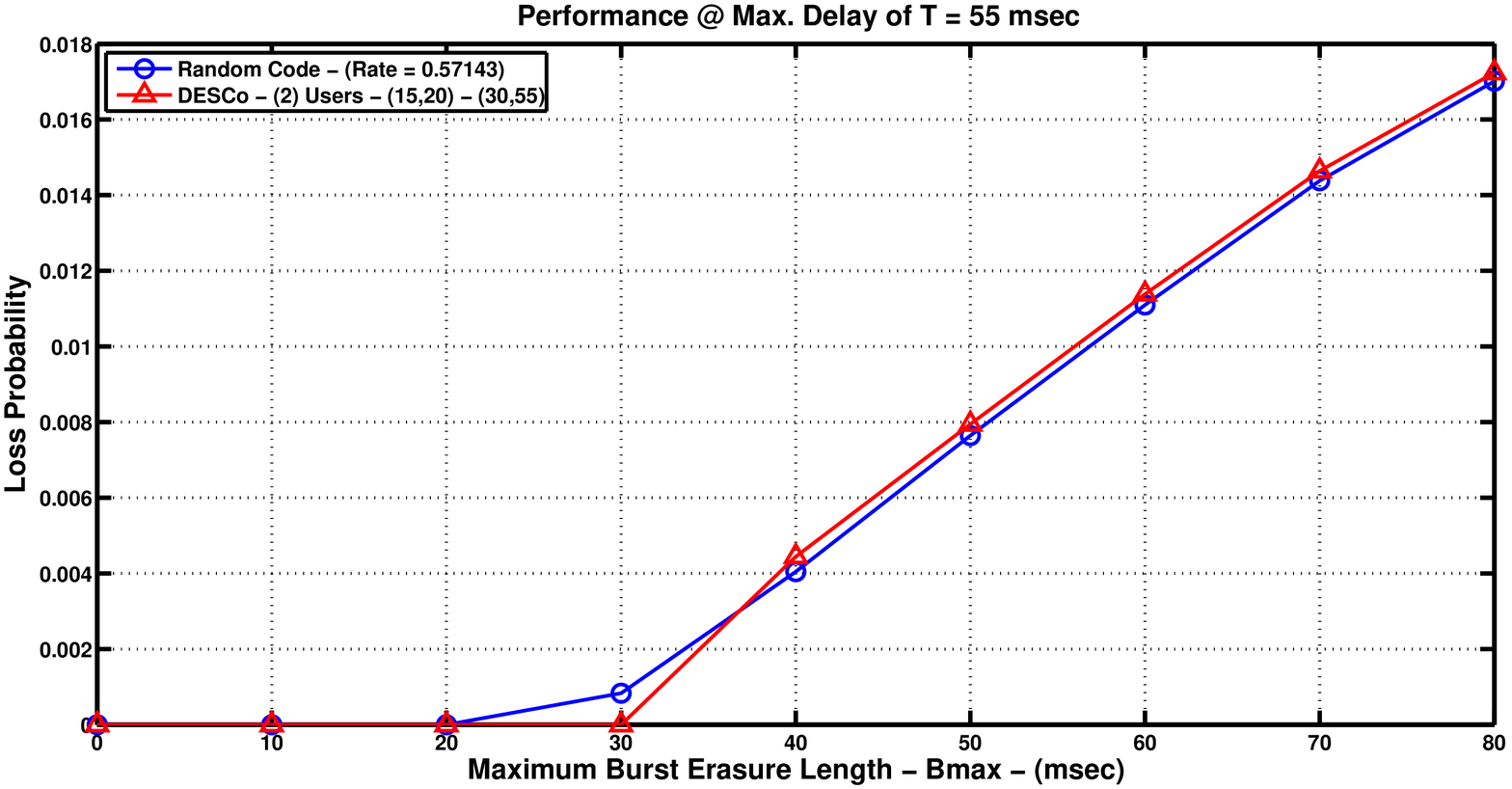}} \\
\resizebox{\columnwidth}{!}{\includegraphics[scale=1]{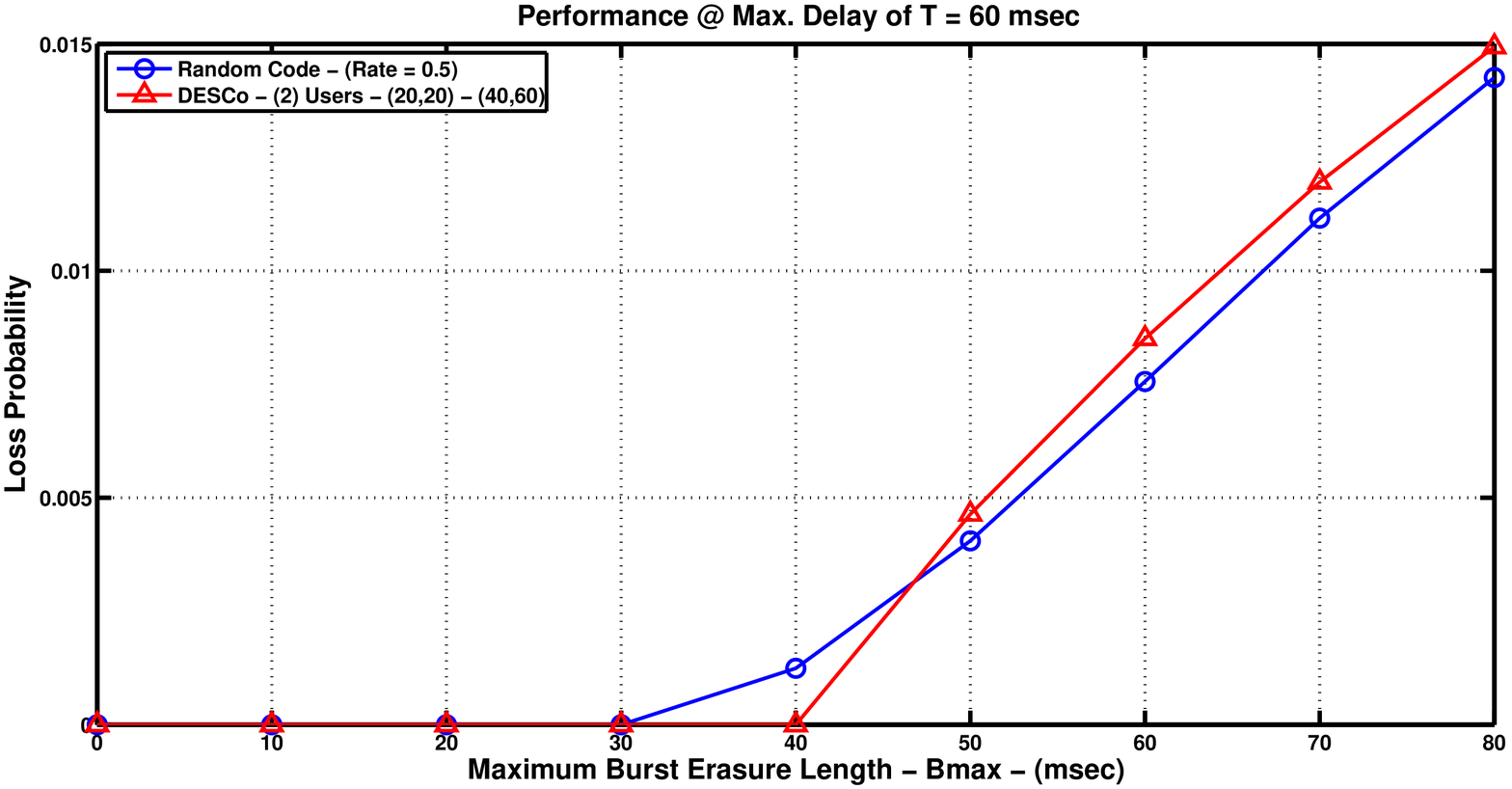}}
\end{tabular}
\caption{Loss Probability at the second receiver.}
\label{NR_User2}
\end{minipage}
\end{figure*}


To examine fundamental performance, we compare between the proposed DE-SCo codes and sequential random linear codes (RLC) numerically and discuss advantages and disadvantages of the proposed codes. The encoder for DE-SCo codes is the one discussed in section~\ref{subsec:DE-SCo_Construction}. For RLC, at each time step $t$ a new source symbol $s[t]$ over an alphabet $\mathcal{S}$ is revealed to the transmitter and encoded into a channel symbol $x[t]$ through a random mapping $f_t(.)$ as follows,
\begin{equation}
x[t] = f_t (s[0],\dots,s[t]),
\end{equation}
i.e., $f_t : \mathcal{S}^t \rightarrow \mathcal{X}$.

In our simulations, we do not construct an explicit function $f_t(\cdot)$ but instead assume that the decoder succeeds with high probability whenever the \emph{instantaneous information debt} becomes non-positive. Intuitively, the information debt is a running sum of the gap between information transmitted over the channel and the information acquired by the receiver. We refer the reader to~\cite[Chapter 9]{Martinian_Thesis},~\cite{MartinianDebt} for details. Our simulated decoder keeps track of the erasure pattern and retrieves the current segment of source symbols as soon as the information debt is non-positive. While every symbol in this setup is ultimately decoded, any symbols that incur a delay that exceeds the maximum delay, are declared to be lost. 

In our simulations we divide the coded data stream into segments of 2000 symbols each and generate one burst erasure in each segment.  Each symbol occupies one millisecond. The burst erasure length is uniformly distributed between $[0,B_{max}]$ symbols and a symbol is declared to be lost if it is not recovered by its deadline. We plot the average loss probability for a stream of $10^5$ segments for both; (1) DE-SCo code with burst-delay parameters $\{ (B,T),(\al B,\al T+B)\}$ for $\al = 2$ and (2) sequential RLC of the same rate for the two users in Fig.~\ref{NR_User1} and Fig.~\ref{NR_User2} respectively as a function of the maximum erasure burst length.

We make a few remarks  on the numerical results. We see that if the maximum size of erasure burst is less than a critical threshold for each scheme then the loss probability is zero. For the DES-Co construction this threshold equals $B_i$. For RLC at rate $R$  it can be easily verified that if the burst-length exceeds $\lceil (1-R)T \rceil$, the first symbol will not be decoded with a delay of $T$.

Next we see that DE-SCo always outperforms RLC for user 1. This can be explained as follows. A rate $R$ DE-SCo can recover completely from an erasure burst of length $B_1$ or smaller for user 1. It fails to recover the erased symbols if the burst length exceeds $B_1$.  The RLC only recovers completely from an erasure burst of length $\lceil (1-R)T_i \rceil$. It provides partial recovery for burst erasures up to length $B_1$ and fails to recover any source symbols when the erasure length exceeds $B_1$. Thus the performance of DE-SCo always dominates RLC for user 1 as illustrated in  Fig.~\ref{NR_User1}.

For user 2, the delay is given by $T_2 = \frac{B_2}{B_1} T_1+ B_1$.  DE-SCo can correct all erasures up to length $B_2$ and fail to recover any symbols if the erasure length is beyond $B_2$. While threshold for perfect recovery for $RLC$ is $\lceil (1-R)T_i \rceil \le B_2$, interestingly it allows for partial recovery for burst lengths up to $\beta = B_2 + \frac{B_1^2}{T_1}$. This threshold is obtained as follows. Suppose an erasure of length $\beta$ occurs at time $t=0,1,\ldots, \beta-1$. The total information debt at this point is $\beta R$. If the information debt becomes non-positive after $\nu$ subsequent non-erased channel symbols then  we must have that $\beta R \le \nu (1-R)$. Substituting $R = \frac{T_1}{T_1+B_1}$ and $\nu = T_2 = \frac{B_2}{B_1} T_1 + B_1$, which is the maximum allowable delay for user 2, we recover the desired threshold.  Since $\beta > B_2$, there is a range of erasure burst lengths where the RLC code can recover a partial subset of source sub-symbols whereas DE-SCo fails to recover any source sub-symbols. This explains why DE-SCo does not outperform random network coding in the high loss regime for user 2.

\section{Conclusion}

This paper constructs a new class of streaming erasure codes that do not commit apriori to a given delay, but rather achieve a delay based on the channel conditions. We model this setup as a multicast problem to two receivers whose channels introduce different erasure-burst lengths and require different delays. The DE-SCo construction embeds new parity checks into the single-user code, in a way such that we do not compromise the single user performance of the stronger user while the supporting the weaker receiver  with an information theoretically optimum delay.  We provide an explicit construction of these codes as well as the associated decoding algorithm. Numerical simulations suggest that these codes outperform simple random linear coding techniques that do not exploit the burst-erasure nature of the channel. 

A number of interesting future directions remain to be explored. The general problem of designing codes that are optimal for any feasible pair $\{(B_1,T_1),(B_2,T_2)\}$ remains open. We expect to report some recent progress along this lines in the near future. While our construction can be naturally extended to more than two users the optimality remains to be seen. Our initial simulation results indicate that the performance gains of the proposed code constructions are limited to burst-erasure channels. Designing codes with similar properties for more general channels remains an interesting future direction.

\vspace{-4em}
\begin{IEEEbiography}[{\includegraphics[width=1in,height=1.25in,clip,keepaspectratio]{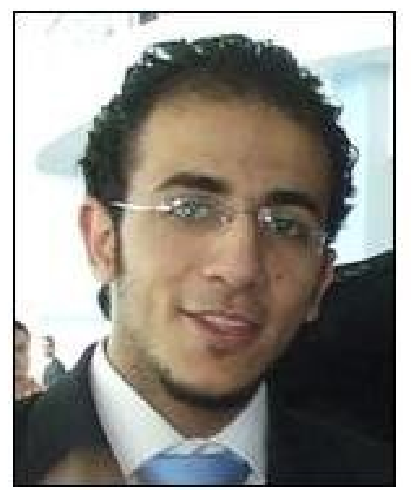}}]{Ahmed Badr}
Ahmed Badr received a B.Sc. degree in Electrical Engineering from Cairo University, Egypt in 2007 and M.Sc. degree in Electrical Engineering from Nile University, Egypt in 2009. From September 2007 to August 2009, he was a Research Assistant in the Wireless Intelligent Networks Center (WINC), Nile University. In September 2009, he assumed his current position as a Research Assistant at Signals Multimedia and Security Laboratory in University of Toronto while pursuing his Ph.D. degree in Electrical Engineering. His research interests include information theory, coding theory and wireless communications.
\end{IEEEbiography}
\vspace{-4em}
\begin{IEEEbiography}[{\includegraphics[width=0.75in,height=1.25in,clip,keepaspectratio]{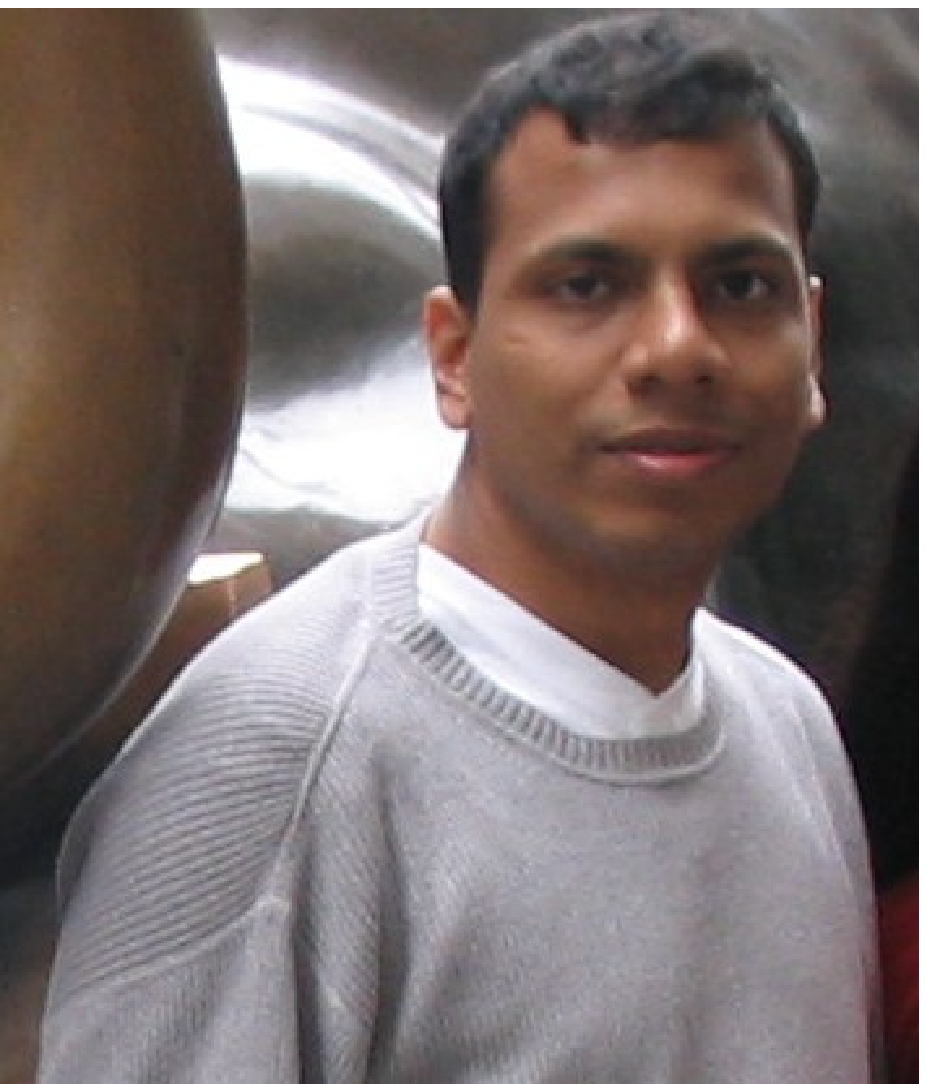}}]{Ashish Khisti}
Ashish Khisti is an assistant professor in the Electrical and Computer Engineering (ECE) department at the University of Toronto, Toronto, Ontario Canada. He received his BASc degree in Engineering Sciences from University of Toronto and his S.M and Ph.D. Degrees from the Massachusetts Institute of Technology (MIT), Cambridge, MA, USA. His research interests span the areas of information theory, wireless physical layer security and streaming in multimedia communication systems. At the University of Toronto, he heads the signals, multimedia and security laboratory. For his graduate studies he was a recipient of the NSERC postgraduate fellowship, HP/MIT alliance fellowship, Harold H. Hazen Teaching award and the Morris Joseph Levin Masterworks award. 
\end{IEEEbiography}
\vspace{-4em}
\begin{IEEEbiography}[{\includegraphics[width=0.75in,height=1.25in,clip,keepaspectratio]{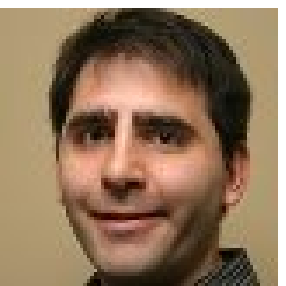}}]{Emin Martinian}
Emin Martinian earned a B.S. from UC Berkeley in 1997, and an S.M and Ph.D. from MIT in 2000 and 2004 (all in electrical engineering and computer science). Emin has been awarded a National Science Foundation Fellowship, the Capocelli Award for Best Paper, second place for best Computer Science Ph.D. at MIT, and over 10 patents. Emin worked at various technology startups (OPC Technologies, PinPoint) and research labs (Bell Labs, Mitsubishi Electric Research Labs) before joining Bain Capital in 2006. He currently works on research and strategy development at Bain's global macro hedge fund.
\end{IEEEbiography}

\end{document}